\documentclass[8pt, showpacs, twocolumn,preprintnumbers,amsmath,amssymb,prb]{revtex4}

\usepackage{graphicx}
\usepackage{dcolumn}
\usepackage{bm}

\begin{document}

\title{Detecting and Switching Magnetization of Stoner Nanograin in Non-local Spin Valve}
\author{Hai-Zhou Lu}
\author{Shun-Qing Shen}
\affiliation{Department of Physics, and Centre of Theoretical and
Computational Physics, The University of Hong Kong, Pokfulam Road,
Hong Kong, China}

\date{\today}

\begin{abstract}
The magnetization detection and switching of an ultrasmall Stoner
nanograin in a non-local spin valve (NLSV) device is studied
theoretically. With the help of the rate equations, a unified
description can be presented on the same footing for the NLSV signal
that reads out the magnetization, and for the switching process. The
setup can be viewed as that the grain is connected to two
non-magnetic leads via sequential tunneling. In one lead, the
chemical potentials for spin-up and -down electrons are split due to
the spin injection in the NLSV. This splitting (or the spin bias) is
crucial to the NLSV signal and the critical condition to the
magnetization switching. By using the standard spin diffusion
equation and parameters from recent NLSV device, the magnitude of
the spin bias is estimated, and found large enough to drive the
magnetization switching of the cobalt nanograin reported in earlier
experiments. A microscopic interpretation of NLSV signal in the
sequential tunneling regime is thereby raised, which show properties
due to the ultrasmall size of the grain. The dynamics at the
reversal point shows that there may be a spin-polarized current
instead of the anticipated pure spin current flowing during the
reversal due to the electron accumulation in the floating lead used
for the readout of NLSV signal.
\end{abstract}


\maketitle


\section{Introduction}

Current-induced magnetization reversal had attracted considerable
interest due to its fundamental significance in understanding
interplay between magnetism and electricity as well as potential
applications in magnetic
memories.\cite{Slonczewski1996JMMM159.L1,Berger1996prb549353,Tsoi1998prl80.4281,Myers1999science285.867,Sun1999.JMMM.202.157,Katine2000.prl.84.3149,Jiang2004.PRL.92.167204}
As the scale of the ferromagnetic nanograins goes down to only
several
nanometers,\cite{Gueron1999.PRL.83.4148,Deshmukh2001.PhysRevLett.87.226801,Jamet2001.prl.86.4676,Thirion2003.natmat.2.524}
many
theoretical\cite{Inoue2004.prb.70.140406R,Braun2004.prb.70.195345,Wetzels2005.prb.72.020407R,Jalil2005.prb.72.214417,Waintal2005prl94_247206,Parcollet2006.prb.73.144420,Fernandez-rossier2007prl98106805,Wang2007.prl.98.077201,Lu2008prb77.235309,Lu2009.prb79.174419}
and experimental works
\cite{Chen2006.prl.96.207203,Krause2007.science.317.1537,Wang2008.apl.93.162501,Garzon2008.prb.78.180401}
were inspired to address the current-induced magnetization reversal
in these small structures. By far, most studied setups were
multi-layer or nanopillar structures with vertical geometries, in
which spins are always carried along the flowing of charge current.
Usually, the critical current density as high as $10^6-10^9$
A/cm$^2$ is required to induce a
reversal.\cite{Ralph2008.JMMM.320.1190} Considering such high
density of current flowing through each nanograin, when a huge
amount of nanograins are integrated in large scale, spurious effects
such as Joule heat, current-induced magnetic field, and noise are
not ignorable.

\begin{figure}[htbp]
\centering
\includegraphics[width=0.45\textwidth]{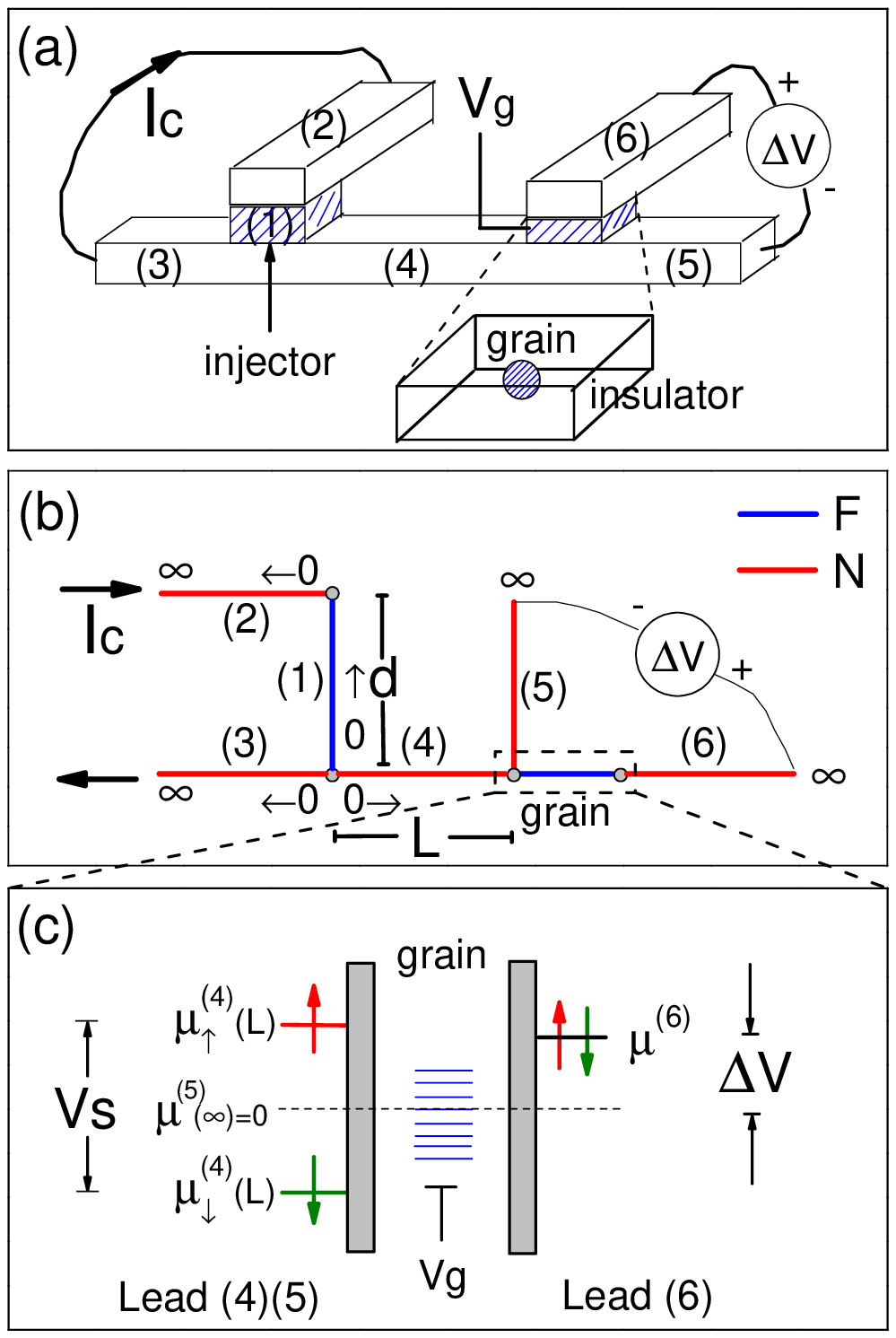}
\caption{(a) The schematic of our non-local spin valve (NLSV)
device. It can be viewed as a combination of the setups employed in
Ref. \onlinecite{Yang2008.NatPhys} and Ref.
\onlinecite{Deshmukh2001.PhysRevLett.87.226801}. (b) The device is
separated into 7 regions. (1) is a fixed ferromagnetic injector of
thickness $d$. The ferromagnetic grain on the right is to be
manipulated and detected. The distance between (1) and the grain is
$L$. `` $0\rightarrow$ " in each region defines the local origin and
positive direction of coordinate. (c) The grain and its nearby
regions is modeled as a Stoner particle coupled to two non-magnetic
lead via tunneling through two barriers. Driven by the
spin-polarized current $I_c$, spins injected from (1) accumulate and
induce a splitting ($V_s$) of the spin-up and down chemical
potentials [$\mu_{\uparrow/\downarrow}^{(4)}(L)$] at the
(4)/(5)/grain interface. $V_g$ is the gate voltage applied to the
insulator [see Fig. 1(a)] surrounding the grain, much like the
technique in Ref. \onlinecite{Deshmukh2001.PhysRevLett.87.226801}.
In this way, it can be used to tune the energy levels inside the
grain with respect to chemical potentials of (4), (5), and (6)
capacitively, while without inducing direct current between the gate
and the grain. $\Delta V$ measures the voltage between (5) and (6),
and is defined as the NLSV voltage. $\mu^{(5)}(\infty)$ is the
chemical potential at the voltmeter side of regions (5), and is set
as the energy zero point throughout the
paper.}\label{fig:setupmodel}
\end{figure}

A possible solution is to use pure spin current, in which the same
amount of spin-up and -down currents flow along opposite directions,
yielding no net electric current. By far, one of the most promising
designs to realize considerable pure spin current is the non-local
spin valve (NLSV)
devices\cite{Johnson1985,Filip2000,Jedema2001,Valenzuela2004,Ji2004,Beckmann2004,Kimura2004,Garzon2005}
with lateral geometry.\cite{Brataas2006.PhysRep.427.157} Recently, a
NLSV was reported to reversibly switch the magnetization of a
ferromagnetic particle.\cite{Kimura2006,Yang2008.NatPhys}  As shown
in Fig. \ref{fig:setupmodel}(a), a typical NLSV includes a bigger
fixed and a smaller free ferromagnets (denoted by the shadow areas)
embedded on the left and right sides, usually referred as injector
and detector, respectively. By driving a current $I_c$ through
regions (2), (1), and (3), spins can be injected from injector (1)
to produce a nonequilibrium spin accumulation in nonmagnetic region
(4). This spin accumulation exhibits as a splitting of chemical
potentials for spin-up and -down electrons (spin bias or spin
voltage), as shown in Figs. \ref{fig:Yang2008_gamma_jle_spinbias}(a)
and \ref{fig:Yang2008_gamma_jle_spinbias}(b). While the
spin-polarized current flows only in the loop formed by regions (2),
(1) and (3), the nonequilibrium spin accumulation in region (4)
diffuses to the right accompanied by a pure spin current flowing
toward the detector. In this way, net charge current is prevented
from flowing directly into the grain. In the response of $I_c$, the
magnetization of the detector can be read out by measuring the
voltage difference $\Delta V $ between (5) and (6), referred to as
the NLSV voltage, which is usually estimated by the spin diffusion
equation.\cite{Son1987,Valet1993,Hershfield1997,Jedema2003PRB,Takahashi2003prb67_052409}
By applying $I_c$ exceeding a critical value, the magnetization of
the detector can be switched
reversibly.\cite{Kimura2006,Yang2008.NatPhys} This process is
usually described by Landau-Lifshitz-Gilbert (LLG)
equation.\cite{Sun2000.prb.62.570}

By far, the detectors are films of size of 100 - 1000 nm.
Considering commercial charge devices already work well at these
scales, the ultimate goal of utilizing pure spin degree of freedom
is to replace the charge devices at only several nanometers.
Besides, because the LLG equation pre-assume the magnitude and
polarization of currents that flow through the grain as input
parameters, the counteraction of the detector on the currents is not
taken into account. As a result, important information could be
missing, e.g., whether the pure spin current is still a pure spin
current after flowing into the detector.

In this work, we study a NLSV device, in which the usual film
detector is replaced with one or multiple well-separated cobalt
nanograins embedded in insulator, as those in Ref.
\onlinecite{Deshmukh2001.PhysRevLett.87.226801}. The nanograin is
much smaller in size ($\sim 10^3$ atoms) than the films and at much
lower temperatures ($\sim 20$ mK). Such small nanograin can be
viewed as a Stoner particle whose ferromagnetism comes from the
exchange interactions between itinerant electrons inside it, and
thereby can be manipulated by exchanging angular momenta with the
electrons that tunnel through
it.\cite{Waintal2003prl91_247201,Waintal2005prl94_247206} The grain
is modeled as coupled to two non-magnetic leads via quantum
tunneling. In one lead, the chemical potentials for the spin-up and
-down electrons are split, due to the spin injection in the NLSV.
The small size of the nanograin allows us to model the NLSV signal
and the magnetization switching within one set of rate equations.
Besides, the previous knowledge of the Co nanograin from
experiments\cite{Gueron1999.PRL.83.4148,Deshmukh2001.PhysRevLett.87.226801,Jamet2001.prl.86.4676,Thirion2003.natmat.2.524}
and theories\cite{Canali2000prl85_5623,Kleff2001prb64_220401} allows
us to perform a realistic evaluation.

This work focuses on two aspects: (i) the possibility of the
detection and switching is evaluated using the parameters extracted
from the previous NLSV\cite{Kimura2006,Yang2008.NatPhys} and Co
grain\cite{Gueron1999.PRL.83.4148,Deshmukh2001.PhysRevLett.87.226801,Jamet2001.prl.86.4676,Thirion2003.natmat.2.524}
experiments. (ii) Such small grain is subjected to strong Coulomb
and magnetic blockades,\cite{Waintal2005prl94_247206} how these
blockades determine the critical conditions for the reversal, e.g.,
the critical driving current $I_c$, the gate voltage $V_g$, and the
spin bias $V_s$.

We find that: (i) The numerical evaluations using realistic
parameters from the recent NLSV\cite{Kimura2006,Yang2008.NatPhys}
and the cobalt grain
experiments\cite{Gueron1999.PRL.83.4148}\cite{Deshmukh2001.PhysRevLett.87.226801,Jamet2001.prl.86.4676,Thirion2003.natmat.2.524}
show that it is possible to employ the NLSV device to detect and
switch the magnetization of a ferromagnetic nanograin under the
present experimental conditions. (ii) Under $I_c$, the NLSV signal
can also be detected in the sequential tunneling regime to read out
the magnetization of the grain, and interpreted from a microscopic
view. Interestingly, if the majority band of the grain is favored to
participate in the electron transport, the sign of the NLSV signal
turns out to be just opposite to that if the the minority band is
preferred to conduct electrons. In the presence of an angle $\theta$
between the easy axis of the grain and the spin-quantization
direction of the lead, the NLSV signal is proportional to $
\cos\theta$ and vanishes when $\theta=\pi/2$. (iii) Under $I_c$
exceeding a critical value, the magnetization of the grain can be
switched reversibly. The critical current $I_c$ required for the
magnetization switching is determined by the gate voltage $V_g$ and
the spin bias $V_s$ at which both the Coulomb and magnetic blockades
in the grain are lifted. By choosing suitable gate voltage, the
critical $V_s$ needed for switching can be minimized to $\sim 2K_N$,
where $K_N$ is the volume-independent anisotropy of the grain.
Besides, the transient current during the magnetization reversal may
be a spin-polarized current instead of the anticipated pure spin
current, due to the electron accumulation or drainage in the
floating lead used for the NLSV measurement. A possible solution is
to remove the floating lead.

The paper is organized as follows. In Sec. \ref{sec:model}, the
model and theoretical formalisms are introduced. In Sec.
\ref{sec:nlsv}, the microscopic NLSV signal in the sequential
tunneling regime is described in detail. In Sec.
\ref{sec:switching}, the magnetization switching under large
injection current $I_c$ is presented. Finally, a summary is given in
Sec. \ref{sec:summary}.

\section{\label{sec:model}Model and theoretical Formalisms}

\subsection{General survey of our setup}

The device we study is shown in Fig. \ref{fig:setupmodel}(a), which
can be divided into 7 regions, ``(1)-(6)" and ``grain", as shown in
Fig. \ref{fig:setupmodel}(b). The different regions of the device
are separated into three parts, and modeled by different formalisms,
depending on their sizes and positions:

(i) the first part consists of regions (1)-(5). Their sizes are
comparable to their spin diffusion lengthes, thus the spin transport
in these regions are governed by the spin \emph{diffusion}
equations.\cite{Son1987,Valet1993,Hershfield1997,Jedema2003PRB,Takahashi2003prb67_052409}
The magnetization of injector (1) is assumed to be fixed.

(ii) the second part is the smaller ( approximately several nm)
grain, which is described as a Stoner
particle,\cite{Canali2000prl85_5623,Kleff2001prb64_220401} coupled
via \emph{sequential tunneling} to two nonmagnetic leads. For
convenience, we call them lead (4,5) and lead (6), respectively.
Lead (4,5) and lead (6) are defined as where regions (4) and  (5)
and region (6) connect the grain, respectively. We assume that there
is no direct tunneling between lead (4,5) and lead (6), i.e., the
only possible connection between them is via the grain.

(iii) The third part is \emph{floating} region (6). Because of the
voltmeter, it is an open circuit between regions (4) and (5) and
region (6), i.e., at steady state, there is no current flowing from
regions (4) and (5) across the grain to (6), due to a voltage
difference between (6) and (5). Throughout the paper, we define the
chemical potential $\mu^{(5)}(\infty)$ at the voltmeter side of
region (5) as the energy zero point. The chemical potential of
region (6) is denoted as $\mu^{(6)}$. The difference between
$\mu^{(6)}$ and $\mu^{(5)}(\infty)$ is denoted as $\Delta
V\equiv\mu^{(6)}-\mu^{(5)}(\infty)$, where $\Delta V$ is the NLSV
voltage. $\Delta V$ will be determined through self-consistent
calculation.

In a word, our model can be viewed as a combination of the setup
used in two kinds of experiments, i.e., the
NLSV\cite{Yang2008.NatPhys} and the transport through ferromagnetic
nanograins.
\cite{Gueron1999.PRL.83.4148,Deshmukh2001.PhysRevLett.87.226801}
Besides:

(i) Different from usual ferromagnetic nanograins
setups,\cite{Gueron1999.PRL.83.4148,Deshmukh2001.PhysRevLett.87.226801}
the chemical potentials for spin-up and -down electrons
$\mu_{\uparrow/\downarrow}^{(4)}(L)$ are split at the (4)/(5)/grain
interface, as shown in Fig. \ref{fig:setupmodel}(c). The splitting
is called the spin bias, and denoted as
$V_s$.\cite{Wang2004.prb.69.205312,Sun2008.prb.77.195313,Lu2008prb77.235309,Lu2009.prb79.174419,Xing2008.apl.93.142107,Stefanucci2008.prb.78.075425}
This spin bias is induced by the spins injected from region (1) to
region (4). Later in Secs. \ref{sec:nlsv} and \ref{sec:criticalIc},
we will see that $V_s$ is crucial to the detection of NLSV signal
and the magnetization reversal, thus we will first evaluate its
magnitude in the following subsection.

(ii) Different for usual NLSVs,\cite{Kimura2006,Yang2008.NatPhys}
The ultrasmall size of the grain allows the NLSV signal, the
magnetization reversal dynamics, and the interaction between the
currents and the grain to be described within one set of rate
equations Eq. (\ref{rateequation}).

\begin{figure}[htbp]
\centering
\includegraphics[width=0.4\textwidth]{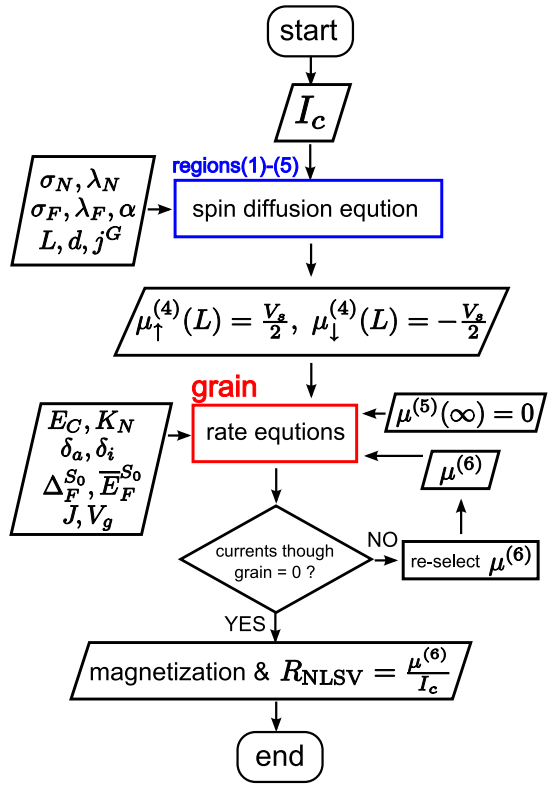}
\caption{The calculation steps of the magnetization of the grain and
the non-local spin valve signal $R_{\mathrm{NLSV}}$ as functions of
$I_c$, the driving current. Rectangles represent calculation
processes. Parallelograms represent inputs or outputs of the
calculations. The region (1), (2), (3), (4) and (5) in Fig.
\ref{fig:setupmodel} are described by the spin diffusion equation
given by Eq. (\ref{diffusionequations}). The grain and its nearby
regions are described by the rate equations given by Eq.
(\ref{rateequation}). The diamond represents the self-consistent
calculation of $\mu^{(6)}$, the chemical potential of region
(6).}\label{fig:procedure}
\end{figure}

For a better understanding of our setup, the calculation steps of
the magnetization and the NLSV signal as functions of $I_c$ is
presented in Fig. \ref{fig:procedure}. This section is organized as
follows. In Sec. \ref{sec:spindiffusion}, $V_s$ will be estimated
with realistic experimental parameters. In Sec.
\ref{sec:grainbrief}, the description of the grain will be
introduced. In Sec. \ref{sec:rateequation}, the transport between
the grain and its leads will be described by the rate equations. The
details of theoretical descriptions can be found in Appendixes
\ref{sec:boundary}-\ref{sec:ValidityRateEquation}, and will be
specified in the following subsections.

\subsection{\label{sec:spindiffusion}Spin diffusion transport in regions (1)-(5) and estimate of spin bias $V_s$}

In this section, we will estimate the magnitude of $V_s$, which is
defined as the splitting of spin-up and -down chemical potentials at
the (4)/(5)/grain interface,
\begin{equation}\label{Vsdefine}
    V_s\equiv\mu_{\uparrow}^{(4)}(L)-\mu_{\downarrow}^{(4)}(L)\equiv\mu_{\uparrow}^{(5)}(0)-\mu_{\downarrow}^{(5)}(0).
\end{equation}
Following the previous literatures, the spin and charge transports
in regions (1)-(5) are described by the standard diffusion
equation,\cite{Son1987,Valet1993,Hershfield1997,Jedema2003PRB,Takahashi2003prb67_052409}
\begin{eqnarray}\label{diffusionequations}
\nabla^2
(\sigma_{\uparrow}\mu_{\uparrow}+\sigma_{\downarrow}\mu_{\downarrow})
&=& 0, \nonumber\\
\nabla^2 (\mu_{\uparrow}-\mu_{\downarrow}) &=&
\frac{1}{\lambda^2}(\mu_{\uparrow}-\mu_{\downarrow}),
\end{eqnarray}
where $\lambda$ is a phenomenological spin-diffusion length that can
be measured experimentally.\cite{Bass2007.jpcm.19.183201} Usually,
the spin-diffusion length in normal metal is much longer than that
in ferromagnetic metal, i.e., $\lambda_N \gg \lambda_F$. In this
work, we use $\lambda_N=1000$ nm, and $\lambda_F=5$
nm.\cite{Yang2008.NatPhys} $\sigma_{\sigma}$ is the conductivity for
$\sigma$ electrons. For ferromagnetic metal,
$\sigma_{\uparrow}=\sigma_F(1+\alpha)/2$ and
$\sigma_{\downarrow}=\sigma_F(1-\alpha)/2$, where $\sigma_F$ is the
total conductivity of the ferromagnetic metal, $\alpha$ is the bulk
polarization defined as $\alpha\equiv
(\sigma_{\uparrow}-\sigma_{\downarrow})/(\sigma_{\uparrow}+\sigma_{\downarrow})$.
For normal metal,
$\sigma_{\uparrow}=\sigma_{\downarrow}=\sigma_N/2$. In this work, we
use $\sigma_N=8.8\times 10^{7}\Omega^{-1}\mathrm{m}^{-1}$,
$\sigma_F=9.8\times 10^{6}\Omega^{-1}\mathrm{m}^{-1}$, and
$\alpha=0.2$.\cite{Kimura2006}

We calculate the chemical potentials in regions (1)-(5) of Fig.
\ref{fig:setupmodel}(b), where regions (2), (3), and (5) can be
regarded as semi-infinite. By assuming that the same cross-section
area $A_{\Box}$ perpendicular to the current in each region, the
solutions to Eq. (\ref{diffusionequations}) can be simplified to be
one dimensional. Following Jedema \emph{et al}.,\cite{Jedema2003PRB}
the general solutions to the 5 regions as functions of position $x$
are given by
\begin{eqnarray}\label{yang_solution}
\mu_{\uparrow/\downarrow}^{(1)} &=&
A+\frac{ej_c}{\sigma_{F}}x\pm\frac{B}{\sigma_{F}(1\pm\alpha)}e^{-\frac{x}{\lambda_{F}}}\pm\frac{C}{\sigma_{F}(1\pm\alpha)}e^{\frac{x}{\lambda_{F}}},\nonumber\\
\mu_{\uparrow/\downarrow}^{(2)}&=&
D+\frac{ej_c}{\sigma_{N}}x\pm\frac{E}{\sigma_{N}}e^{-x/\lambda_{N}}, \nonumber\\
\mu_{\uparrow/\downarrow}^{(3)}&=&
-\frac{ej_c}{\sigma_{N}}x\pm\frac{F}{\sigma_{N}}e^{-x/\lambda_{N}},\nonumber\\
\mu_{\uparrow/\downarrow}^{(4)}&=&
\pm\frac{G}{\sigma_{N}}e^{-x/\lambda_{\mathrm{N}}}\pm\frac{H}{\sigma_{N}}e^{x/\lambda_{N}},\nonumber\\
\mu_{\uparrow/\downarrow}^{(5)}&=&
\pm\frac{K}{\sigma_{N}}e^{-x/\lambda_{\mathrm{N}}},
\end{eqnarray}
where the origin and positive direction in each region are locally
defined for a concise form of the general solutions, and indicated
by `` $0\rightarrow$ " in Fig. \ref{fig:setupmodel}(b).
$j_c=I_c/A_{\Box}$, where $I_c$ is the spin-polarized total current
flowing through regions (2), (1), and (3). $e$ is the electron
charge. The 9 coefficients $A$, $B$, $C$, $D$, $E$, $F$, $G$, $H$,
and $K$ will be determined by boundary conditions.

The boundary conditions are given as follows:

(i) The current polarization usually loses when flowing through an
interface from the ferromagnetic to the normal side, due to, e.g.,
the spin-dependent scattering. We take this loss into account by
phenomenologically introducing an efficiency parameter $\gamma
\in[0,1]$,
\begin{eqnarray}\label{efficiencyparameter}
 \gamma\frac{j_{F\uparrow}-j_{F\downarrow}}{j_{F\uparrow}+j_{F\downarrow}}
=
\frac{j_{N\uparrow}-j_{N\downarrow}}{j_{N\uparrow}+j_{N\downarrow}},
\end{eqnarray}
where $j_{\sigma}=-(\sigma_{\sigma}/e)\partial_x\mu_{\sigma}$, is
the current density for the spin-$\sigma$ electrons. By combining
Eq. (\ref{efficiencyparameter}) and the conservation of the total
current
\begin{eqnarray}\label{totalconserve}
j_{F\uparrow}+j_{F\downarrow}= j_{N\uparrow}+j_{N\downarrow},
\end{eqnarray}
the boundary conditions at the ferromagnetic/normal interface are
then given by,
\begin{eqnarray}
j_{F\uparrow} =
\frac{\gamma+1}{2\gamma}j_{N\uparrow}+\frac{\gamma-1}{2\gamma}j_{N\downarrow},\nonumber\\
 j_{F\downarrow} =
\frac{\gamma+1}{2\gamma}j_{N\downarrow}+\frac{\gamma-1}{2\gamma}j_{N\uparrow}.
\end{eqnarray}

(ii) The chemical potentials of each spin components are continuous
at each interface. For simplicity, we do not explicitly include the
spin-dependent chemical potential drops caused by the interface
resistance. Its destructive effects, in particular on the reduction
in the spin bias, will be approximately accounted by considering a
relatively small injection efficiency $\gamma$.

(iii) At the (4)/(5)/grain interface, the spin-up and -down currents
flowing into the grain are $I^G$ and $-I^G$, respectively, i.e.,
(how this boundary condition is derived can be found in Appendix
\ref{sec:boundary})
\begin{eqnarray}
-A_{\Box}\frac{\sigma_N}{2}\partial_x\mu_{\uparrow/\downarrow}^{(4)}(L)
=\pm
eI^G-A_{\Box}\frac{\sigma_N}{2}\partial_x\mu_{\uparrow/\downarrow}^{(5)}(0),
\end{eqnarray}
where we assume a pure spin current flowing from (4) and (5) into
the grain, which however may not be true as we will see in Sec.
\ref{sec:purespincurrent}. However, this deviation is neglected
because $I^G$ is too small to affect $V_s$, as we will see in Sec.
\ref{sec:validjg}. Actually, because $I^G$ is ignorably small, we
simply neglect it in our numerical calculations, though we include
it in equations explicitly.

\begin{figure}[htbp]
\centering\includegraphics[width=0.5\textwidth]{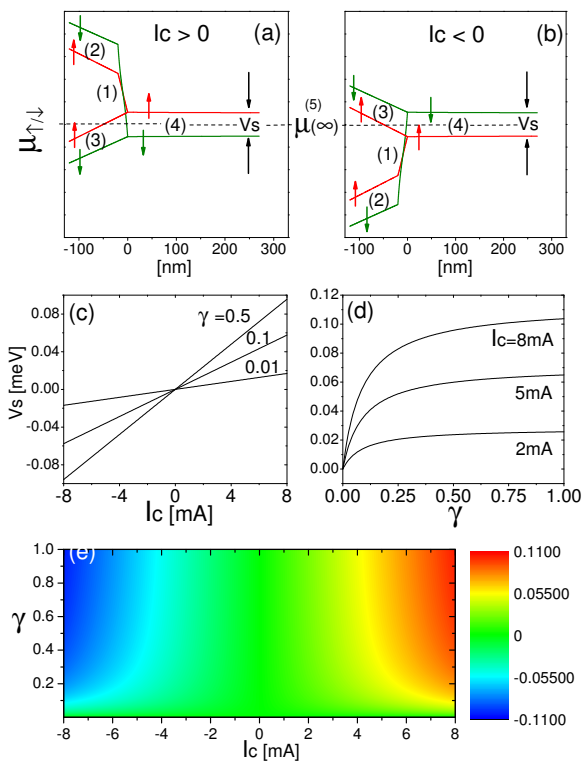}
\caption{(a)(b): The chemical potentials $\mu_{\uparrow}$ (marked by
$\uparrow$) and $\mu_{\downarrow}$ (marked by $\downarrow$) in
regions (1)-(4). They change with the direction of $I_c$, the
spin-polarized current flowing through regions (1), (2), and (3).
$V_s$ is defined as $\mu_{\uparrow}-\mu_{\downarrow}$ at the
(4)/(5)/grain interface. The horizontal dashed line indicates
$\mu^{(5)}(\infty)$, the energy zero point throughout the paper. (c)
$V_s$ vs $I_c$ for different values of spin injection efficiency
$\gamma$. (d) $V_s$ vs $\gamma$ for different $I_c$. (e) $V_s$ (in
unit of meV) as a function of $I_c$ and $\gamma$. The parameters for
(c)(d)(e): $d=20$ nm, $L=270$ nm, $A_{\Box}=170\times65\
\mathrm{nm}^2$, $\lambda_F=5$ nm, $\lambda_N=1000$ nm,
$\sigma_F=9.8\times 10^6\ \Omega^{-1}\mathrm{m}^{-1}$,
$\sigma_N=8.8\times 10^7\ \Omega^{-1}\mathrm{m}^{-1}$, $\alpha=0.2$,
$I^G=0$.} \label{fig:Yang2008_gamma_jle_spinbias}
\end{figure}

With these boundary conditions, the coefficients from $A$ through
$K$ in Eq. (\ref{yang_solution}) are readily found. Figs.
\ref{fig:Yang2008_gamma_jle_spinbias}(a) and
\ref{fig:Yang2008_gamma_jle_spinbias}(b) show the spin-resolved
chemical potentials $\mu_{\uparrow}$ and $\mu_{\downarrow}$ in
regions (1)-(5). For a clear demonstration of the splitting between
spin-up and -down chemical potentials, we choose $\alpha=0.9$ only
in these two figures. For the realistic evaluations in the rest part
of this work, we choose $\alpha=0.2$, as estimated by the
experiments.\cite{Kimura2006,Yang2008.NatPhys} All the
conductivities, spin-diffusion lengthes, and the bulk polarization
used for the evaluation are extracted from the experiment
data,\cite{Kimura2006,Yang2008.NatPhys} and given in Fig.
\ref{fig:Yang2008_gamma_jle_spinbias}. Driven by $I_c$, the spins
injected from injector (1) induce a splitting between
$\mu_{\uparrow}$ and $\mu_{\downarrow}$ at interfaces (1)/(2) and
(1)/(3)/(4). Because injector (1) is sandwiched between two normal
metals, $\mu_{\uparrow}$ and $\mu_{\downarrow}$ in (1) cross with
each other at the center and split oppositely on the opposite sides.
In each region of (1), (2), and (3), because of the charge current
$I_c$, $\mu_{\uparrow}$ and $\mu_{\downarrow}$ not only split but
also demonstrate steep slopes with the same trend. Because of no net
charge current, $\mu_{\uparrow}$ and $\mu_{\downarrow}$ in region
(4) just diffuse to the right side in opposite gradients (not
obvious here because $L\ll \lambda_N$).

By using Eq. (\ref{Vsdefine}), the analytic expression for $V_s$ is
found out as
\begin{eqnarray}\label{Vs}
V_s&\equiv &\mu_{\uparrow}^{(4)}(L)-\mu_{\downarrow}^{(4)}(L)\equiv\mu_{\uparrow}^{(5)}(0)-\mu_{\downarrow}^{(5)}(0),\nonumber\\
&=&2ej_c\frac{\lambda_F}{\sigma_F}\frac{\alpha}{(1-\alpha^2)}\frac{(M_1-2)}{M_2}e^{-\frac{L}{\lambda_N}}
\nonumber\\
&-&2e\frac{I^G}{A_{\Box}}[\frac{\lambda_F}{\sigma_F}\frac{2e^{-\frac{2L}{\lambda_N}}M_1}{\gamma(1-\alpha^2)
M_2}
    +\frac{\lambda_N}{\sigma_N}(1-e^{-\frac{2L}{\lambda_N}})],
\end{eqnarray}
where
\begin{eqnarray}
M_1 &=&2\cosh(\frac{d}{\lambda_F})+2M_0\sinh(\frac{d}{\lambda_F}),     \nonumber\\
M_2
&=&6M_0\cosh(\frac{d}{\lambda_F})+2(2M_0^2+1)\sinh(\frac{d}{\lambda_F}),\nonumber\\
M_0 &=&
\frac{\sigma_N\lambda_F}{\sigma_F\lambda_N\gamma(1-\alpha^2)}.
\end{eqnarray}
Figures. \ref{fig:Yang2008_gamma_jle_spinbias}(c)-(e) show the spin
bias $V_s$ as a function of $I_c$ and $\gamma$. Later we will see in
Secs. \ref{sec:purespincurrent} and \ref{sec:validjg} that $I^G$ is
ignorably small, thus hardly changing $V_s$. Thus, we let $I^G=0$.
As we see, $V_s$ can be as large as $\sim 0.1$ meV for the present
device and parameters. $V_s$ increases linearly with $I_c$, while
logarithmically with $\gamma$. $V_s$ changes sign with the current
$I_c$.

In the following, for a conservative and realistic simulation, we
will always choose a relatively low injection efficiency
$\gamma=0.1$ and assume $I^G=0$. Since the above boundary conditions
only consider the conservation of current density, our results are
valid when assuming the identical cross-section area in each part of
the setup. One should consider different cross section area in
different regions and current conservation for a more general
case.\cite{Takahashi2003prb67_052409}

\subsection{\label{sec:grainbrief}Many-body states of the ferromagnetic nanograin}

In this work, we will describe the ferromagnetic nanograin by using
the minimal possible
model\cite{Canali2000prl85_5623,Kleff2001prb64_220401} proposed to
describe the experiment transport spectra through cobalt
nanograin.\cite{Gueron1999.PRL.83.4148,Deshmukh2001.PhysRevLett.87.226801}
This model was also adopted to discuss spin-polarized
current-induced relaxation and spin torque in the ferromagnetic
nanograin.\cite{Waintal2003prl91_247201,Waintal2005prl94_247206,Parcollet2006.prb.73.144420}

A full and detailed description of this model can be found in
Appendix \ref{sec:graintheory}. Simply speaking, at low
temperatures, the grain can be described by the many-body states
$|N, S,S_z\rangle$, where $N$ is the total electron number inside
the grain, $S$ and $S_z$ are the magnitude and the $z$-component of
the total angular momentum $\mathbf{S}$ of the grain, respectively.

In the following, we will focus on two branches of states. The first
branch is
\begin{eqnarray}\label{branch1}
|N_0, S_0,S_z\rangle,
\end{eqnarray}
i.e., there are $N=N_0$ electrons inside the grain, and the
magnitude of the total angular momentum of the grain is $S=S_0$.
Besides, since $S_z\in[-S_0,S_0]$, there are $2S_0+1$ states in this
branch. The second branch is obtained by adding an extra electron to
the minority band of the grain with respect to the first branch, so
that the total electron number increases by 1 and the magnitude of
the total angular momentum decreases by $1/2$. This branch is
denoted as
\begin{eqnarray}\label{branch2}
|N_0+1, S_0-1/2,S_z\rangle,
\end{eqnarray}
where $S_z\in[-S_0+1/2,S_0-1/2]$, i.e., there are $2S_0$ states in
this branch.

We refer regions (4)-(6) as the two ``leads" connected to the grain,
one is from regions (4) and (5), the other is from region (6). By
tuning the gate voltage $V_g$ and applying the spin bias $V_s$
(induced by $I_c$), the energies of the two branches presented in
Eqs. (\ref{branch1}) and (\ref{branch2}) can be set to be nearly
degenerate with respect to the chemical potentials
$\mu^{(4)}_{\uparrow,\downarrow}(L)$,
$\mu^{(5)}_{\uparrow,\downarrow}(0)$, and $\mu^{(6)}$. In this
situation, the electrons in lead (4,5) and lead (6) can be exchanged
with the grain. Then we can use the polarization of the exchanged
electrons to detect and manipulate the magnetization of the
ferromagnetic grain.

\begin{figure}[htbp]
\centering
\includegraphics[width=0.45\textwidth]{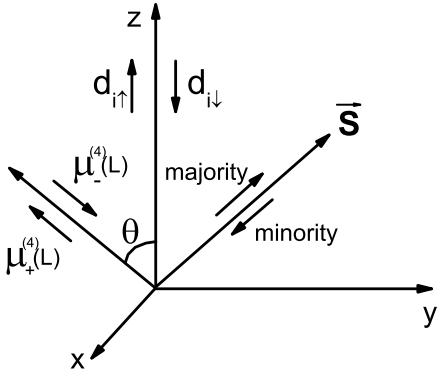}
\caption{The easy-axis of the grain is set as $z$ direction. The
grain electrons ($d_{i\uparrow/\downarrow}$) are quantized along $z$
direction. The majority and minority electrons orient parallel and
anti-parallel with $\mathbf{S}$, respectively. $\mathbf{S}$ is the
total angular momentum of the grain formed by all the electrons
inside the grain. $\theta$ is the angle between the easy axis of
grain and the spin-quantization direction of lead
(4,5).}\label{fig:axis}
\end{figure}

\subsection{\label{sec:rateequation}Rate equations in the presence of
spin bias }

The evolution of the many-body states of the grain by exchanging
electrons with the weakly coupled lead (4,5) and lead (6) is
described by the Pauli rate
equations.\cite{Beenakker1991.PRB.44.1646,Blum1996book,vonDelft2001.PhysRep.345.61}
We only consider the sequential tunneling regime. Born approximation
and Markoff approximation are applied, and $H_{\mathrm{T}}$ is
treated by perturbation up to the second order.\cite{Blum1996book}
The rate equation can be expressed in a compact form,
\begin{eqnarray}\label{rateequation}
\partial _{t}P_{l}=\sum_{l'}R_{ll'}P_{l'},
\end{eqnarray}
where $0 \leq P_{l} \leq 1$ are the probability to find the state
$l\equiv |N,S,S_z\rangle$. The diagonal and off-diagonal terms of
the coefficient matrix of the rate equations are, respectively,
\begin{eqnarray}
R_{l'\neq l} = \sum_{\alpha\sigma i }R_{l'\neq l}^{\alpha\sigma i},\
\ R_{ll}=-\sum_{l'\neq l} R_{l'l},
\end{eqnarray}
where
\begin{eqnarray}\label{ratecoefficient}
R_{l'\neq l}^{\alpha \uparrow i} &=&
\Gamma_{\alpha}\cos^2\frac{\theta_{\alpha}}{2} [|\langle
l'|d_{i\uparrow }|l\rangle
|^{2}f(E_{l}-E_{l'}-\mu_{+ }^{\alpha}) \nonumber\\
&&\ \ \ \ \ \ \ \ \ \ \  +|\langle l| d_{i\uparrow }|l'\rangle
|^{2}f(E_{l}-E_{l'}+\mu_{+}^{\alpha})]\nonumber\\
&&+ \Gamma_{\alpha}\sin^2\frac{\theta_{\alpha}}{2} [|\langle
l'|d_{i\uparrow }|l\rangle
|^{2}f(E_{l}-E_{l'}-\mu_{-}^{\alpha}) \nonumber\\
&&\ \ \ \ \ \ \ \ \ \ \ +|\langle l| d_{i\uparrow }|l'\rangle
|^{2}f(E_{l}-E_{l'}+\mu_{-}^{\alpha})],
\end{eqnarray}
and one just replaces $\uparrow$ with $ \downarrow $ and exchanges
$+ $ and $-$ to obtain $R_{l'l}^{\alpha\downarrow i}$. Note that the
Fermi distribution $f(x)=1/[\exp (x/k_{B}T)+1]$ is spin-resolved.
The parameter $\Gamma_{\alpha} =2\pi \sum_{k}|V_{k\alpha i
}|^{2}\delta (\omega -\epsilon _{k\alpha})$ represents the
spin-irrelevant coupling between lead $\alpha\in\{(4,5),(6)\}$ and
the grain. For simplicity, we assume that
$\Gamma_{(4,5)}=\Gamma_{(6)}=\Gamma$, and $\Gamma$ are assumed to be
independent of the specific single-particle level $i$. The
overlapping $\langle l|d_{i\sigma }|l'\rangle$ can be found by
calculating the Clebsch-Gordan coefficients.

$\theta_{\alpha}$ in Eq. (\ref{ratecoefficient}) is the angle
between the easy axis of grain and the spin-quantization direction
of lead $\alpha$. For simplicity, we set the easy-axis of the grain
as $z$-axis and assume that $\theta_{(6)}=0$ and
$\theta_{(4,5)}=\theta\in[0,\pi/2]$, as shown in Fig.
\ref{fig:axis}. In the following, we denote
$\mu_{+/-}^{(4)}=\mu_{\uparrow/\downarrow}^{(4)}$ when $\theta=0$.

In terms of $P_{l}$, the magnetization of the grain is given by
\begin{eqnarray}
M=\sum_{l}S_z^{l}P_{l},
\end{eqnarray}
and the spin-$\sigma$ current flowing from region $\alpha$ into the
grain are defined as
\begin{eqnarray}
I_{\alpha}^{\sigma}=-e
\sum_{ll'}(N_{l}-N_{l'})R_{l'l}^{\alpha\sigma}P_{l'},
\end{eqnarray}
where $N_{l}$ and $S_z^{l}$ correspond to the $N$ and $S_z$ of the
state $l\equiv |N,S,S_z\rangle$.

The validity of the rate equations is discussed in Appendix
\ref{sec:ValidityRateEquation}.

\section{\label{sec:nlsv}NLSV signal in the tunneling regime}

In this section, we will present the microscopic picture of the NLSV
signal in the sequential tunneling regime. When a small current
$I_c$ is driven in the loop formed by regions (1)-(3) of Fig.
\ref{fig:setupmodel}(a), the relative alignment of magnetization
between the two ferromagnets can be read out by measuring the
voltage difference between regions (6) and (5).

\subsection{$\theta=0 $ case}

\begin{figure}[htbp]
\centering
\includegraphics[width=0.5\textwidth]{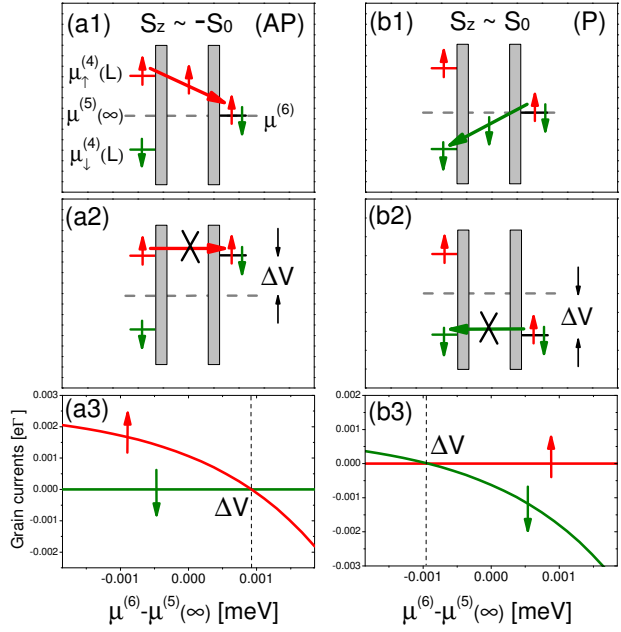}
\caption{Using NLSV signal in the sequential tunneling regime to
read out the magnetization of the grain. (a1)-(a3) and (b1)-(b3)
show the cases when the magnetization of the grain is antiparallel
(AP) and parallel (P) with injector (1), respectively.
$\mu^{(5)}(\infty)$ and $\mu^{(6)}$ are the chemical potentials on
the voltmeter side of regions (5) and (6), respectively. (a1) When
$S_z\sim -S_0$, the grain favors spin-up current flowing from lead
(4,5) to lead (6). Because it is an open circuit between lead (4,5)
and lead (6), the electrons flowing into lead (6) can not go
anywhere but build up and raise the chemical potential $\mu^{(6)}$.
(a2) Once the chemical potential $\mu^{(6)}$ is raised to be aligned
with $\mu^{(4)}_{\uparrow}(L)$, the current flowing and electron
building-up stop. Then one can measure a voltage difference between
$\mu^{(5)}(\infty)$ and $\mu^{(6)}$, which corresponds to the NLSV
voltage. (a3) The spin-up and -down currents as functions of
$\mu^{(6)}-\mu^{(5)}(\infty)$. The NLSV voltage $\Delta
V=\mu^{(6)}-\mu^{(5)}(\infty)$ corresponds to the vertical dashed
line at which both currents are zero. (b1)-(b3) can be understood
similarly. The parameters: $\theta=0$, $I_c=250\ \mu$A,
$\gamma=0.1$, $I^G=0$, $S_0=100$, $T=20$ mK, $\Delta V_g=-K_N$, and
other parameters are the same as those in Fig.
\ref{fig:Yang2008_gamma_jle_spinbias}. }\label{fig:nlsv}
\end{figure}

We will discuss first the case when $\theta=0$. For $\theta=0$, we
denote $\mu_{+/-}^{(4)}=\mu_{\uparrow/\downarrow}^{(4)}$. The
results can be easily generalized for $\theta\neq0$ in Sec.
\ref{sec:thetaneq0}. The left and right columns of Fig.
\ref{fig:nlsv} show the cases when the magnetization of the grain is
anti-parallel (AP) and parallel (P) with injector (1), respectively.
First, we consider the AP case, i.e., $S_z$ of the grain $\sim
-S_0$. By putting the experimental measurement
current\cite{Yang2008.NatPhys} $I_c=$250 $\mu$A into Eq. (\ref{Vs})
and assuming $\gamma=0.1$ and $I^G=0$, we estimate that a positive
spin bias $V_s\sim$ 1.84 $\mu$eV will be induced at the interface
where regions (4), (5) and the grain connect, so that
$\mu^{(4)}_{\uparrow/\downarrow}(L)=\mu^{(5)}_{\uparrow/\downarrow}(0)=\pm
V_s/2$. It is natural to assume that the distance between the
voltmeter and the (4)/(5)/grain interface $\gg \lambda_N =1\ \mu$m,
so the split chemical potentials for spin-up and -down electrons
will decay to only one chemical potential $\mu^{(5)}(\infty)$ on the
voltmeter side of region (5). Because region (5) is nonmagnetic, the
decays of spin-up and -down chemical potentials are symmetric, so it
can be anticipated that
$\mu^{(5)}(\infty)=[\mu^{(4)}_{\uparrow}(L)+\mu^{(4)}_{\downarrow}(L)]/2=[\mu^{(5)}_{\uparrow}(0)+\mu^{(5)}_{\downarrow}(0)]/2$.
This situation is depicted by $\mu^{(4)}_{\uparrow/\downarrow}(L)$
and $\mu^{(5)}(\infty)$ in Fig. \ref{fig:nlsv} (a1).

In the response of this small spin bias, a small current will be
generated, flowing through the grain. Specifically, spin-up current
will be favored for the AP case. This can be understood with the
help of the Clebsch-Gordan coefficient,
\begin{eqnarray}\label{CGcoefficient}
&& \langle N_0,S_0,S_z|d_{i\uparrow}| N_0+1,
S_0-\frac{1}{2},S_z+\frac{1}{2}\rangle \nonumber\\
&=& \langle
j_1=S_0,m_1=S_z;j_2=\frac{1}{2},m_2=\frac{1}{2}\nonumber\\
&&|J=S_0-\frac{1}{2},M=S_z+\frac{1}{2}
\rangle \nonumber\\
&=&- \sqrt{\frac{S_0-S_z}{2S_0+1}}\nonumber\\
&&\langle N_0,S_0,S_z|d_{i\downarrow}|
N_0+1,S_0-\frac{1}{2},S_z-\frac{1}{2}\rangle\nonumber\\
&=& \langle
j_1=S_0,m_1=S_z;j_2=\frac{1}{2},m_2=-\frac{1}{2}\nonumber\\
&&|J=S_0-\frac{1}{2},M=S_z-\frac{1}{2}
\rangle \nonumber\\
&=&\sqrt{\frac{S_0+S_z}{2S_0+1}},
\end{eqnarray}
apparently, for $S_z\sim -S_0$, the probability for spin-up
electrons to tunnel through the grain
\begin{eqnarray}\label{CGcoefficientup}
\frac{S_0-(-S_0)}{2S_0+1} \sim  1,
\end{eqnarray}
is much larger than the probability for spin-down electrons
\begin{eqnarray}\label{CGcoefficientdown}
\frac{S_0+(-S_0)}{2S_0+1} \sim  0.
\end{eqnarray}
In other words, the spin-down current is magnetic
blockaded.\cite{Waintal2005prl94_247206} Note that the spin
selection rules remain qualitatively unchanged even for small
fluctuation of $S_z$ around $-S_0$, as long as $S_0\gg1$.

As a result, the favored spin-up electrons will flow from region (4)
through the grain into region (6), if
$\mu^{(4)}_{\uparrow}(L)>\mu^{(6)}$. Remember that there is a
voltmeter between regions (5) and (6), so the electrons tunneling
into region (6) can not go anywhere but accumulate and raise the
chemical potential of region (6) until $\mu^{(6)}\sim
\mu^{(4)}_{\uparrow}(L)$. After this accumulation is accomplished,
no more current will flow and there is finally a stable voltage
difference $\Delta V$ between $\mu^{(6)}$ and $\mu^{(5)}(\infty)$,
as shown in Fig. \ref{fig:nlsv} (a2).

Fig. \ref{fig:nlsv} (a3) shows how to numerically determine the NLSV
voltage $\Delta V$. One just scan $\mu^{(6)}$ and calculate the
tunneling current through the grain. $\Delta V$ is then found out as
at which $\mu^{(6)}-\mu^{(5)}(\infty)$ both the spin-up and -down
currents vanish, as indicated by the vertical dashed line. It turns
out that in the sequential tunneling regime and when $\theta=0$, the
magnitude of the NLSV voltage is
\begin{eqnarray}
|\Delta V^{\theta=0}| &=& |V_s|/2.
\end{eqnarray}
By using Eq. (\ref{Vs}) and the parameters given in Fig.
\ref{fig:nlsv}, for $I_c=250\ \mu$A, the NLSV voltage $\Delta
V^{\theta=0}$ is found out to be about 0.92 $\mu$V. The NLSV signal
is thereby equal to
\begin{equation}\label{nlsv}
    R_{\mathrm{NLSV}}^{\theta=0}\equiv \frac{\Delta V^{\theta=0}}{I_c} = \frac{0.92\ \mu\mathrm{V}}{250 \mu \mathrm{A}}\simeq
    3.7
    \mathrm{m} \Omega.
\end{equation}

Similarly, as shown in Figs. \ref{fig:nlsv} (b1)-(b3), the parallel
(P) case favors that spin-down current flowing from region (6)
through the grain to regions (4) and (5), also due to the same spin
selection rules Eq. (\ref{CGcoefficient}). This current will drain
the electrons in region (6) and lower $\mu^{(6)}$ until
$\mu^{(6)}\sim \mu^{(4)}_{\downarrow}(L)$. Therefore, for the P
case, $\Delta V^{\theta=0}=-0.92\mu$eV and
$R^{\theta=0}_{\mathrm{NLSV}}=-3.7$m$\Omega$, right opposite to the
AP case.

\subsection{\label{sec:thetaneq0}$\theta\neq 0 $ case}

\begin{figure}[htbp]
\centering
\includegraphics[width=0.45\textwidth]{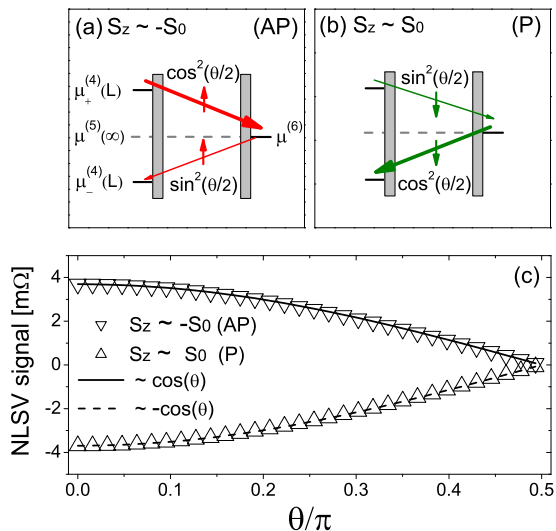}
\caption{NLSV signal when $\theta\neq 0$. (a) when $\theta\neq 0$
and the magnetization of the grain is $S_z=-S_0$ (AP), the spin-up
current can flow simultaneously from lead (4,5) to lead (6) with
relative probability $\cos^2\frac{\theta}{2}$, and from lead (6) to
lead (4,5) with probability $\sin^2\frac{\theta}{2}$. (b) when
$\theta\neq 0$ and the magnetization of the grain is $S_z=S_0$ (P),
spin-down current can flow simultaneously from lead (4,5) to lead
(6) with relative probability $\sin^2\frac{\theta}{2}$, and from
lead (6) to lead (4,5) with probability $\cos^2\frac{\theta}{2}$.
(c) The NLSV signal as a function of $\theta$ for the P
($\triangle$) and AP ($\nabla$) cases. $\pm \cos\theta$ are also
plotted for comparison. Other parameters are the same as Fig.
\ref{fig:nlsv}. }\label{fig:nlsvtheta}
\end{figure}

When $\theta\neq0$, the chemical potentials of lead (4,5) now are
denoted as $\mu_{+}^{(4)}(L)$ and $\mu_{-}^{(4)}(L)$, respectively.
We still take $S_z\sim-S_0$ for example. As shown in Fig.
\ref{fig:nlsvtheta}(a), when $\theta\neq0$, the favored
spin-$\uparrow$ current can flow not only from $\mu_{+}^{(4)}(L)$ to
$\mu^{(6)}$, but also from $\mu^{(6)}$ to $\mu_{-}^{(4)}(L)$,
according to the rate equations Eq. (\ref{ratecoefficient}). These
two currents are denoted as $I^{\uparrow}_{\mu_{+}^{(4)}(L)
\rightarrow \mu^{(6)}}$ and $I^{\uparrow}_{ \mu^{(6)}
\rightarrow\mu_{-}^{(4)}(L)} $, respectively.

According to the rate coefficients Eq. (\ref{ratecoefficient}), the
spin-up electrons are related to $\mu^{(4)}_{+}(L)$ by
$\cos^2\frac{\theta}{2}$, and to $\mu^{(4)}_{-}(L)$ by
$\sin^2\frac{\theta}{2}$. Besides, lead (6) is not spin-dependent,
so the polarization and magnitude of the current between the grain
and (6) just follow those between (4) and the grain. Therefore,
\begin{eqnarray}
I^{\uparrow}_{\mu_{+}^{(4)}(L) \rightarrow \mu^{(6)}} &\propto&
\cos^2\frac{\theta}{2},\nonumber\\
I^{\uparrow}_{ \mu^{(6)} \rightarrow\mu_{-}^{(4)}(L)} &\propto&
\sin^2\frac{\theta}{2},
\end{eqnarray}
respectively. As a result, $\mu^{(6)}$ will saturate to a balanced
position at which the above two opposite currents cancel with each
other. Because
$\cos^2\frac{\theta}{2}-\sin^2\frac{\theta}{2}=\cos\theta$, this
balanced position of $\mu^{(6)}$ turns out to be proportional to
$\cos\theta$.\cite{Kimura2007.prl.98.156601}

The triangles in Fig. \ref{fig:nlsvtheta}(c) show the
self-consistent results of the NLSV signal as a function of
$\theta$. For comparison, the functions $\pm \cos\theta$ are also
plotted by solid and dashed lines, respectively. One easily
concludes that the NLSV signal in the presence of $\theta$ is given
by
\begin{eqnarray}
R_{\mathrm{NLSV}} = \pm \frac{V_s}{2I_c}\cos\theta,
\end{eqnarray}
where $\pm $ depends on the magnetization of the grain.

\subsection{Discussion}

There are several points should be clarified:

(i) $R_{\mathrm{NLSV}}$ we obtained for the present device and
parameters happen to be of the same order as the experimental
observations, where the NLSV signals are of the order of $1\sim 10$
m$\Omega$.\cite{Yang2008.NatPhys}

(ii) Above, we only consider the case that $S$ decreases by $1/2$
when the extra electron is added, where $S$ is the magnitude of the
total angular momentum of the grain. Also, according to Fig.
\ref{fig:phasediagram}, there is small probability that $S$
increases by $1/2$ when adding the extra electron, i.e., the extra
electron is preferred to be added to the majority band, then the
spin selection rules will become totally reversed as
\begin{eqnarray}
\langle N_0,S_0,S_z
|d_{i\uparrow}|N_0+1,S_0+\frac{1}{2},S_z+{\frac{1}{2}}\rangle&=&
\sqrt{\frac{S_0+S_z+1}{2S_0+1}},\nonumber\\
\langle N_0,S_0,S_z
|d_{i\downarrow}|N_0+1,S_0+\frac{1}{2},S_z-{\frac{1}{2}}\rangle &=&
\sqrt{\frac{S_0-S_z+1}{2S_0+1}},\nonumber\\
\end{eqnarray}
so that tunneling of spin-up (spin-down) electrons will be favored
when $S_z\sim S_0$ ($S_z\sim -S_0$). Therefore, the results will be
totally reversed. This is a direct consequence of the strong Coulomb
repulsion and the unequal spacings $\delta_a$ and $\delta_i$ of
majority and minority one-particle levels for an ultrasmall grain,
and a major difference from the relatively large
films.\cite{Kimura2006,Yang2008.NatPhys}

(iii) Note that in regions (4) and (5), the stead-state splitting of
spin-up and -down chemical potentials is maintained by the spins
continuously injected by $I_c$. This is different from in region
(6), where there should be no current flowing in or out at steady
state, because of the voltmeter. Therefore, in region (6), the
spin-up and -down electrons will eventually relax to one chemical
potential for sufficient long time. That is why we consider, for the
steady-state solution, only one spin-irrelevant chemical potential
$\mu^{(6)}$ in region (6).

(iv) In the simulation, although the parameters listed in Table
\ref{tab:parameters} for a $S_0=1000$ grain are exploited, we only
use $S_0=100$ for simulation because of the limited computing power.
We have checked the results from $S_0=10$ through $S_0=100$, and the
results turn out to be quantitatively unchanged as long as $S\gg 1$.

\section{\label{sec:switching}Magnetization Switching}

In this section, we will present the magnetization switching of the
grain under a current $I_c$ exceeding a critical value. This
critical $I_c$ is determined by the gate voltage $V_g$ and the spin
bias $V_s$, at which the strong Coulomb and magnetic blockades of
the grain are lifted. Then currents can flow through the grain, and
transfer the angular momentum carried by the flowing electrons to
the grain. This will be discussed in Sec. \ref{sec:criticalIc}.

Besides, still because of these blockades, one can not pre-assume
there is always a pure spin current flowing through the grain, so
whether the reversal is accompanied by the pure spin current is
waited to be checked. This will be discussed in Sec.
\ref{sec:purespincurrent}.

\subsection{\label{sec:criticalIc}Critical $I_c$ and $V_s$}

\begin{figure}[htbp]
\centering
\includegraphics[width=0.5\textwidth]{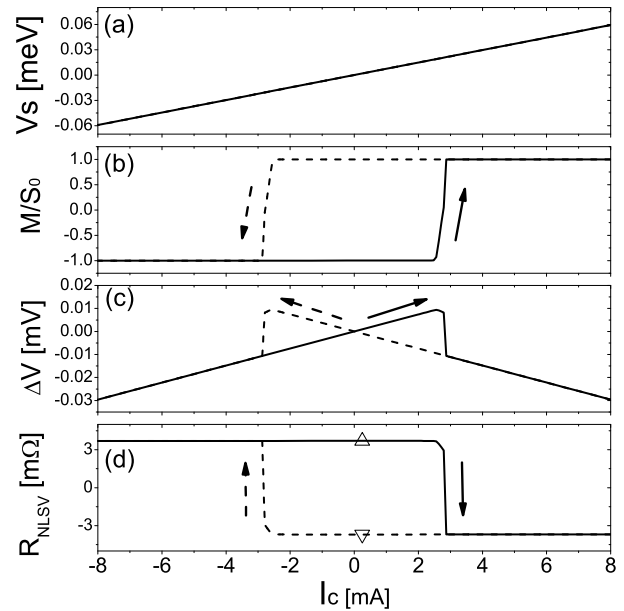}
\caption{(a) The spin bias $V_s$, (b) the normalized magnetization,
(c) the NLSV voltage, and (d) the NLSV signal as functions of the
driving current $I_c$. The parameters $\theta=0$, $\gamma=0.1$,
$I^G=0$, $S_0=100$, $T=20$ mK, $\Delta V_g=0$, and other parameters
are the same as those in Fig. \ref{fig:Yang2008_gamma_jle_spinbias}.
}\label{fig:hysteresis}
\end{figure}

As we have discussed, the electron tunneling between the lead and
the grain are subjected to the Coulomb and magnetic blockades
simultaneously. According to Eq. (\ref{chargingenergy_Vg}), by
choosing suitable gate voltage, the charging energy $E_C$ can be
compensated, but the transition energies still depend on the
magnetization of the grain, i.e., the grain may be magnetic
blockaded.\cite{Waintal2005prl94_247206} In this situation, this
magnetic blockade can be lifted by applying the spin bias $V_s$
exceeding a critical value, which thereby defines the minimal
critical $I_c$ and $V_s$.

We will use the reversal from $S_z=-S_0$ to $S_0$ to extract the
minimal critical $I_c$ and $V_s $ at which the magnetic blockade can
be lifted and the switching can be performed.

Suppose the grain is initially prepared at the state $|N_0, S_0,
-S_0\rangle$. By adding a spin-up electron from lead (4,5) into the
minority band of the grain, the grain will transit to the state
$|N_0+1, S_0- \frac{1}{2}, -S_0+\frac{1}{2}\rangle$. This
transition, energetically requires that
\begin{equation}
    \mu_{\uparrow}^{(4)}(L)> E_{|N_0+1, S_0- \frac{1}{2},-S_0+\frac{1}{2}\rangle}-E_{|N_0,
    S_0,-S_0\rangle}.
\end{equation}
Via this transition, $S_z$ increases by $1/2$ unit.

Further, by draining a spin-down electron from the minority band of
the grain to lead (4,5), the grain will transit from the state
$|N_0+1, S_0-\frac{1}{2},-S_0+\frac{1}{2}\rangle$ to $|N_0, S_0,
-S_0+1\rangle$. This energetically requires that
\begin{equation}
    \mu_{\downarrow}^{(4)}(L)< E_{|N_0+1, S_0- \frac{1}{2},-S_0+\frac{1}{2}\rangle}-E_{|N_0,
    S_0,-S_0+1\rangle}.
\end{equation}
Via this transition, $S_z$ also increases by 1/2 unit.

Note that $\mu_{\uparrow/\downarrow}^{(4)}(L)=\pm V_s/2$. If one
applied a sufficient large $I_c$, so that the spin bias $V_s$ driven
by $I_c$ is large enough, for all the possible $S_z\in [-S_0, S_0]$,
there are always
\begin{eqnarray}\label{ntop}
\mu_{\downarrow}^{(4)}(L)<\{ E_{|N_0+1,S_0-
\frac{1}{2},S_z\pm\frac{1}{2}\rangle}-E_{|N_0,S_0,S_z\rangle}\}<\mu_{\uparrow}^{(4)}(L).\nonumber\\
\end{eqnarray}
Then, one can expect a sequence of consecutive charging-discharging
steps to be driven, which charges the grain with only spin-up
electrons and discharges the grain with only spin-down electrons. As
a result of this charging-discharging sequence, the magnetization of
the grain will eventually be reversed from $S_z=-S_0$ to $S_0$.

Similarly, to reverse the magnetization from $S_z=S_0$ to $-S_0$,
the energy requirement is that for all the possible $S_z$,
\begin{eqnarray}\label{pton}
\mu_{\uparrow}^{(4)}(L)<\{ E_{|N_0+1,S_0-\frac{1}{2},S_z\pm
\frac{1}{2}\rangle}-E_{|N_0,S_0,S_z\rangle}\}<\mu_{\downarrow}^{(4)}(L).\nonumber\\
\end{eqnarray}

Therefore, the minimal required $V_s$, which equals
$\mu_{\uparrow}^{(4)}(L)-\mu_{\downarrow}^{(4)}(L)$, is determined
by the width of spectrum $E_{|N_0+1,S_0-\frac{1}{2},S_z\pm
\frac{1}{2}\rangle}-E_{|N_0,S_0,S_z\rangle}$ for all
$S_z\in[-S_0,S_0]$. According to Eq. (\ref{chargingenergy_Vg}), this
spectrum width is $2K_N$, which thereby set the value for the
minimal required spin bias. The minimal critical current is then
defined as the $I_c$ by which the generated $V_s\geq 2K_N$.

Fig. \ref{fig:hysteresis} shows $V_s$, magnetization, and NLSV
signals as functions of $I_c$, when $\theta=0$. For each $I_c$,
$V_s$ is calculated first using Eq. (\ref{Vs}). Then, $V_s$ is put
into the rate equations Eq. (\ref{rateequation}) to
self-consistently determine $\Delta V$ and $\mu^{(6)}$ by using the
same method shown in Figs. \ref{fig:nlsv} (a3) and (b3) until both
spin-up and -down currents through the grain vanish. Finally, the
magnetization is obtained by putting the calculated $\mu^{(6)}$ back
to the rate equations. The NLSV signal $R_{\mathrm{NLSV}}$ is found
by $R_{\mathrm{NLSV}}=\Delta V/I_c$. These steps is shown by Fig.
\ref{fig:procedure}.

The triangles $\nabla$ and $\triangle$ in Fig. \ref{fig:hysteresis}
indicate the P and AP cases we have already discussed in Fig.
\ref{fig:nlsv}, respectively. Keep increasing $I_c$ until $V_s$
exceeds $2K_N$, the magnetization of the grain will be reversed. In
the present set of parameters, the steady-state magnetization starts
to reverse when $|I_c|$ is a little larger than 2.5 mA. The
switching is accomplished after $|I_c|$ exceeds $\sim 3$ mA, at
which $|V_s|$ is right larger than $2K_N=0.02$ meV. We attribute the
broadening of reversal point at $|I_c|=2.5\sim 3$ mA to the thermal
fluctuation of the lead electron bath. At the reversal point, both
the sign and slope of $\Delta V$ changes abruptly.
$R_{\mathrm{NLSV}}$ also demonstrates a hysteresis loop in analogy
to the hysteresis loop of the magnetization, but with opposite
signs. This opposition has already been explained in the discussion
(ii) of Sec. \ref{sec:nlsv}.

There are several points needed to be clarified:

(i) The results should be qualitatively unaffected for small
$\theta\neq 0$, because $\theta$ does not change
$E_{|N_0+1,S_0-\frac{1}{2},S_z\pm
\frac{1}{2}\rangle}-E_{|N_0,S_0,S_z\rangle}$, while only these
energy differences determine the critical spin bias $V_s$ and $I_c$.

(ii) We have concluded that the minimal critical current is only
related to $K_N$, which does not depend on $S_0$, so we use
$S_0=100$ to perform the calculation. We have checked that the
simulation results for other $S_0\gg 1$ turn out to be qualitatively
unchanged.

(iii) According to Eq. (\ref{chargingenergymin}),
$E_{|N_0+1,S_0-\frac{1}{2},S_z\pm
\frac{1}{2}\rangle}-E_{|N_0,S_0,S_z\rangle}$ is also a function of
the gate voltage $V_g$, therefore the critical $I_c$ can be tuned by
the gate voltage. Above we only discuss the minimal critical $I_c$,
i.e., the case when the Coulomb and the one-particle energies are
already compensated by choosing suitable gate voltage (details can
be found in Appendix \ref{sec:stateforsimulation}). Therefore, the
spin bias only has to lift the magnetic blockade and thereby can be
minimized. If the nearly degenerate situation were tuned away by a
magnitude of $\Delta V_g$, the spin bias $V_s$ then has to
compensate the Coulomb and magnetic blockades simultaneously. Then,
the critical $V_s$ will become $|2K_N|+|\Delta V_g|$, which thereby
requires larger critical $I_c$.

(iv) The critical current density. For the present parameters, the
critical current is about $I_c=$ 3 mA, while the cross-section area
is $A_{\Box}=170\times 65\ \mathrm{nm}^2$. Therefore, the critical
current density is about
\begin{eqnarray}
\frac{I_c}{A_{\Box}}=\frac{3\times 10^{-3}\mathrm{A}}{170\times
10^{-7}\mathrm{cm}\times 65\times 10^{-7}\mathrm{cm}}&\approx & 2.7
\times 10^7 \mathrm{A/cm}^2
\end{eqnarray}
This value is comparable to most experiments of nanopillars and
multi-layers.\cite{Ralph2008.JMMM.320.1190}

\subsection{\label{sec:purespincurrent}Pure spin current ?}

\begin{figure}[htbp]
\centering
\includegraphics[width=0.5\textwidth]{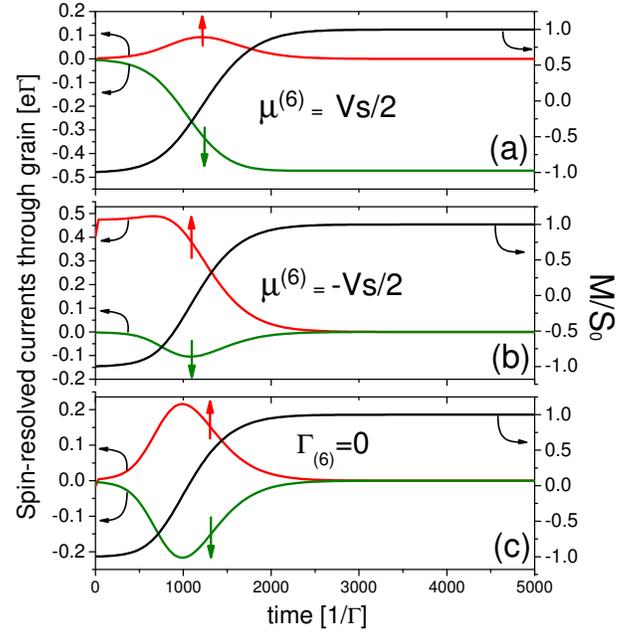}
\caption{The spin-up and -down currents flowing through the grain
and the magnetization of the grain as functions of time when
assuming that the chemical potential of region (6) is: (a)
$\mu^{(6)}=V_s/2$, (b) $\mu^{(6)}=-V_s/2$. (c) region (6) is not
connected to the grain. The positive direction of current is defined
as flowing from lead (4,5) to lead (6). $\theta=0$, $V_s=3K_N$,
$\gamma=0.1$, $T=20$ mK, $S_0=100$, $\Delta V_g=0$, and other
parameters are the same as those in Fig.
\ref{fig:Yang2008_gamma_jle_spinbias}. }\label{fig:purecurrent}
\end{figure}

We are particularly interested in whether there is truly a pure
current flowing through the grain during the reversal in the present
device. Therefore, we studied a situation that the grain is
initialized at $S_z=-S_0$ and $I_c$ is suddenly switched on to
generate a spin bias large enough to drive a switching from
$S_z=-S_0$ to $S_0$, then see how the spin-resolved currents and the
magnetization evolve with time.

Roughly speaking, if we assume the electrons tunneling through the
grain transfer all their angular momenta to the grain, to reverse
the grain from $S_z=-S_0$ to $S_0$, there should be at least $4S_0$
electrons tunneling through the grain during the reversal. Because
the grain is much smaller than region (6), it is safe to expect that
the electrons flowing in and out region (6) along with the reversal
process will hardly change $\mu^{(6)}$.

According to Fig. \ref{fig:nlsv} (a2), when $V_s$ is positive while
$S_z=-S_0$, $\mu^{(6)}$ will saturate at $V_s/2$; and when $V_s>0$,
$S_z=S_0$, $\mu^{(6)}$ will saturate at $-V_s/2$. So in the
following we will compare two limits. The first limit assumes that
$\mu^{(6)}=V_s/2$. The second assumes that $\mu^{(6)}=-V_s/2$. These
two limits are shown in Figs. \ref{fig:purecurrent} (a) and (b),
respectively.

Let us first consider the first limit $\mu^{(6)}=V_s/2=0.015$ meV.
This situation is the same as shown in Fig. \ref{fig:nlsv} (a2), but
$V_s$ is much larger in magnitude. According to Fig.
\ref{fig:purecurrent}(a), as the magnetization changes from $-S_0$
to $S_0$, the spin-down current gradually becomes favored and the
spin-up current unfavored, which is consistent with the spin
selection rules Eq. (\ref{CGcoefficient}). Remember the
configuration of chemical potentials remains unchanged during the
reversal as shown in Fig. \ref{fig:nlsv} (a2). As a result, the
magnitude of spin-down current will keep growing during the
reversal, and even after the magnetization is reversed to $S_0$,
there will still be a steady spin-down current flowing from lead (6)
to lead (4,5). Although not shown in our result, one can expect, for
sufficient long time, the spin-down current leaking from region (6)
will eventually shift $\mu^{(6)}$ down to $-V_s/2$, much like the
same situation as shown by Fig. \ref{fig:nlsv} (b2). Then the
spin-down current will cease to flow.

The second limit is shown by Fig. \ref{fig:purecurrent} (b). At the
start, $S_z=-S_0$, so spin-up current is favored to flow from lead
(4,5) to lead (6). As the reversal from $-S_0$ to $S_0$ goes on, the
spin-up current will gradually become unfavored, and drop to zero
after the reversal is accomplished. Although the spin-down current
is favored when $S_z=S_0$, the Fermi levels for spin-down electrons
on both sides of the grain are the same, so neither the spin-up nor
-down current will be flowing after the reversal.

After considering the above two limits, it is natural to expect that
a real situation should be between them, i.e., $\mu^{(6)}$ should
float gradually from $V_s/2$ to $-V_s/2$, and there should be a
spin-polarized current instead of the anticipated pure spin current
flowing through the grain during and after the reversal, until the
accumulation or drainage in floating region (6) is accomplished.

The extra charge part of the current is produced along with the
electron accumulation or drainage process in region (6). To prove
this point, we decouple region (6) completely by letting
$\Gamma_{(6)}=0$.\cite{Lu2009.prb79.174419} The results are shown in
Fig. \ref{fig:purecurrent} (c). In this case, the spin-up and down
currents are always the same in magnitude and opposite in direction,
i.e., there is only a pure spin current flowing. Besides, by
integrating the current over time using the data of Fig.
\ref{fig:purecurrent}(c),
\begin{eqnarray}
Q_{\uparrow/\downarrow}=\int_0^{\infty}
I^{\uparrow/\downarrow}_{(4,5)}\ dt,
\end{eqnarray}
we obtain that both the spin-up electrons that tunnel into the grain
and spin-down electrons off the grain are $2S_0$. This is consistent
with the change of $S_z$ by $2S_0$ during the reversal, where half
is from the $2S_0$ incoming spin-up electrons and half the $2S_0$
outgoing spin-down electrons.

\subsection{\label{sec:validjg}Validity of approximation $I^G=0$}

Because surrounded by insulator, the grain is connected to the leads
via the quantum tunneling. The reference 20 of Ref.
\onlinecite{Kleff2001prb64_220401} estimates that the tunneling rate
for Ref. \onlinecite{Gueron1999.PRL.83.4148} is around $\Gamma =
10^9$ $\mathrm{s}^{-1}$. According to Fig. \ref{fig:hysteresis},
where either the magnitude of spin-up or -down current is smaller
than $e\Gamma$. Therefore, we estimate the current in and out the
grain $I^G$ are well smaller than
\begin{eqnarray}
e\Gamma = 1.6\times 10^{-19} \mathrm{Coulomb}\times 10^{9}s^{-1} =
1.6 \times 10^{-10}A
\end{eqnarray}
This value is 6 - 7 orders smaller than, e.g., the driven current
$I_c$ ($\sim 10^{-3}$A) shown in our Fig. \ref{fig:hysteresis}. So
it is valid for us to ignore $I^G$ in our
numerical calculations, though we explicitly kept it in Eq. \ref{Vs}.\\

\section{\label{sec:summary}Summary  }

In this work, we theoretically studied the magnetization switching
and detection of a ferromagnetic nanograin in a non-local spin valve
(NLSV) device.

Different from the original experiment,\cite{Yang2008.NatPhys} our
nanograin is much smaller in size and at much lower temperatures,
thus subjected to strong Coulomb and magnetic blockades. We describe
the grain as a Stoner particle, whose ferromagnetism comes from the
exchange interactions between itinerant electrons inside it. Because
of the ultrasmall size, the one-particle levels inside it are
quantized. Because of the ferromagnetism, the level spacings for the
majority and minority electrons are unequal.

As shown in Fig. \ref{fig:setupmodel}(a), the grain is coupled to
regions (4)-(6) of the NLSV device via quantum tunneling. Regions
(4) and (5), and region (6) can be regarded as two nonmagnetic
leads, respectively. In the lead formed by regions (4) and (5), a
spin-dependent splitting of chemical potentials (spin bias) is
induced by the spin-polarized current $I_c$ injected from
ferromagnetic injector (1), as shown in Figs.
\ref{fig:Yang2008_gamma_jle_spinbias}(a) and
\ref{fig:Yang2008_gamma_jle_spinbias}(b). The other electrode (6) is
floating and connected to region (5) through a voltmeter.

By applying a $I_c$ and measure the voltage difference $\Delta V$
between regions (5) and (6), the magnetization of the grain can be
read out by the NLSV signal $R_{\mathrm{NLSV}}=\Delta V/I_c$.
Because of the unequal level spacings for the majority and minority
electrons and Coulomb blockade, the NLSV signal in the tunneling
regime depends not only on the magnetization of the grain, but also
on whether the majority or minority band of the grain is favored to
contribute to the electron transport. The results when the minority
band is favored are right opposite to when the majority band is
favored. In the presence of an angle $\theta$ between the easy-axis
of the grain and the spin-quantization direction of the electrode,
the NLSV signal is proportional to $\cos\theta$ and vanishes when
$\theta=\pi/2$.

By applying $I_c$ exceeding a critical value, the magnetization of
the grain can be switched reversibly by the spin bias generated by
the $I_c$. Because of the strong Coulomb and magnetic blockades, the
electron flowing between the grain and the electrodes is not
possible unless both blockades are lifted, then the angular momenta
carried by the flowing electrons can be transferred to the grain.
Therefore, the critical value of $I_c$ to drive the magnetization
switching is determined by: at what gate voltage $V_g$ and spin bias
$V_s$, both the Coulomb and magnetic blockades can be lifted. We
also show that the current accompanying the switching may not be a
pure spin current, due to the accumulation or drainage of electrons
in the floating lead used for the NLSV measurement. A possible
solution is to remove the floating lead.

Our numerical evaluations using realistic parameters from the recent
NLSV\cite{Kimura2006,Yang2008.NatPhys} and the cobalt grain
experiments\cite{Gueron1999.PRL.83.4148,Deshmukh2001.PhysRevLett.87.226801,Jamet2001.prl.86.4676,Thirion2003.natmat.2.524}
show that it is possible to employ the NLSV device to detect and
switch the magnetization of a ferromagnetic nanograin under the
present experimental conditions.

\section{Acknowledgements}

We thank Wei-Qiang Chen, Zhong-Yi Lu, Rong Lv, Chao-Xing Liu,
Zhan-Feng Jiang, Rui-Lin Chu, and Wen-Yu Shan for helpful
discussions. This work was supported by the Research Grant Council
of Hong Kong under Grant No. HKU 704809 and HKU 10/CRF/08.

\appendix

\section{\label{sec:boundary}Boundary condition (iii)}

\begin{figure}[htbp]
\centering\includegraphics[width=0.25\textwidth]{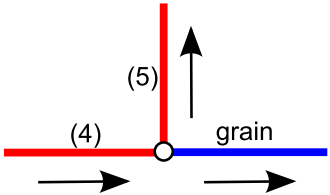}
\caption{The zoom-in of Fig. \ref{fig:setupmodel}(b) near interface
(4)/(5)/grain. Arrows mark the locally defined positive direction in
each region. Three regions meet at the central node. In the local
coordinates of (4), the node is at $L$. In the local coordinates of
(5), the node is at $0$.} \label{fig:boundary3}
\end{figure}

The boundary condition can be found with the help of Fig.
\ref{fig:boundary3}, which is a zoom-in of Fig. 1(b) near interface
(4)/(5)/grain. The positive direction of the locally defined
coordinate in each section is marked by arrow. With the help of
these arrows, one can write
\begin{eqnarray}
I^{(4)}_{\sigma} &=& I^G_{\sigma}+I^{(5)}_{\sigma},
\end{eqnarray}
i.e., the current flowing into the node equals to those flowing out.
The current density is related to electrochemical potential by
$j_{\sigma}=-(\sigma_{\sigma}/e)\partial_x \mu_{\sigma}$. We will
take the spin-up component as a example. In (4) and (5),
$\sigma_{\uparrow}=\frac{\sigma_N}{2}$; and the spin-up current
flowing from the node to the grain is defined as
$I^G_{\uparrow}\equiv I^G$. Put these together, one can easily
obtain that
\begin{eqnarray}
-A_{\Box}\frac{\sigma_N}{2e}\partial_x
\mu^{(4)}_{\uparrow}(L)=I^G-A_{\Box}\frac{\sigma_N}{2e}\partial_x
\mu^{(5)}_{\uparrow}(0).
\end{eqnarray}
Similarly, for spin-down $I^G_{\downarrow}\equiv -I^G$, so
\begin{eqnarray}
-A_{\Box}\frac{\sigma_N}{2e}\partial_x
\mu^{(4)}_{\downarrow}(L)=-I^G-A_{\Box}\frac{\sigma_N}{2e}\partial_x
\mu^{(5)}_{\downarrow}(0)
\end{eqnarray}

\section{\label{sec:graintheory}The quantum theory of the ferromagnetic nanograin}

\subsection{\label{sec:grainmodel}Model of ferromagnetic nanograin}

In this work, we will describe the ferromagnetic nanograin by using
the minimal possible
model\cite{Canali2000prl85_5623,Kleff2001prb64_220401} proposed to
describe the experiment transport spectra through cobalt
nanograin.\cite{Gueron1999.PRL.83.4148,Deshmukh2001.PhysRevLett.87.226801}
This model was then adopted to discuss spin-polarized current
induced relaxation and spin
torque.\cite{Waintal2003prl91_247201,Waintal2005prl94_247206,Parcollet2006.prb.73.144420}
It also provided a starting point to study the Kondo resonance in
the STM spectrum of a ferromagnetic cluster on metal
surface.\cite{Fiete2002.PRB.66.024431} A more detailed microscopic
tight-binding model with exchange interactions and atomic spin-orbit
couplings was also
proposed,\cite{MacDonald2001.SolidStateComm.119.253,Cehovin2002.PRB.66.094430,Cehovin2003.PRB.68.014423}
to reveal a unified origin of the magnetic anisotropy as well as
collective and quasiparticle excitations in the ferromagnetic
nanograin. Besides, the Jaynes-Cummings model also reproduced the
transport features by considering the interaction between
particle-hole excitation and
magnon.\cite{Michalak2006.prl.97.096804}

In this work, we focus on how the collective spin and one-particle
excitations of the grain react to the spin bias, therefore the
minimal model is adequate for the current topic. The Hamiltonian for
the grain and its couplings to nearby leads is given by
\begin{eqnarray}
H_{\mathrm{total}}=H_{G}+H_{\mathrm{lead}}+H_{\mathrm{T}},
\end{eqnarray}
where the Hamiltonian for the ferromagnetic grain takes the form,
\begin{eqnarray}\label{HG}
H_{G} =
\sum_{i\sigma}\epsilon_{i\sigma}d_{i\sigma}^{\dag}d_{i\sigma}-\frac{J}{2}\mathbf{\hat{S}}\cdot
\mathbf{\hat{S}}-\frac{K_N}{S_0} \hat{S}_z^2+E_C \delta
\hat{N}^2+V_g\delta
\hat{N}\nonumber\\
\end{eqnarray}
where the first term stands for the kinetic energy of electrons in
the grain, $d_{i\sigma}$ ($d_{i\sigma}^{\dag}$) annihilates
(creates) an electron on the one-particle level $i$ in the grain,
with energy $\epsilon_{i\sigma}$ and spin
$\sigma=\{\uparrow,\downarrow\}$. $\mathbf{\hat{S}}=\sum_i
\frac{1}{2}\sum_{\sigma\sigma'}d_{i\sigma}^{\dag}\overrightarrow{\tau}_{\sigma\sigma'}d_{i\sigma'}$
is the total angular momentum of the grain electrons,
$\overrightarrow{\tau}$ is the vector of Pauli matrices. $J$ is a
phenomenological constant depicting the exchange interactions
between each pair of electrons in the grain. $\hat{S}_z$ is the
$z$-component of $\mathbf{\hat{S}}$. $K_N$ is the volume-independent
anisotropic constant. In this work, we consider that the fluctuation
of the electron number ($\sim 1$) is much smaller than than the
itinerant electrons $>1000$ already in the grain, so the fluctuation
of $K_N$ as a function of the electron number is ignored. $\delta N$
is the number of extra electrons added into the grain, compared with
a reference electron number $N_0$ already in the grain. $E_C$ is the
charging energy required to add the excess electrons in the grain,
which we will see later can be compensated by applying a gate
voltage $V_g$.

We refer the part where regions (4), (5), and (6) connect the grain
as two ``leads", one is from regions (4) and (5) together, and the
other from region (6). For convenience, we call them lead (4,5) and
lead (6), respectively. The Hamiltonian for the leads takes the form
\begin{equation}
H_{\mathrm{lead}}=\sum_{k ,\alpha,\tau }\epsilon_{k
\alpha}c_{k\alpha \tau }^{\dag }c_{k \alpha\tau },
\end{equation}
where $c_{k\alpha \tau }^{\dag }(c_{k\alpha \tau })$ is the creation
(annihilation) operator for a continuous state in lead $\alpha\in
\{(4,5),(6)\}$ with energy $\epsilon_{k \alpha}$ and spin
$\tau\in\{+,-\}$. The tunneling between the grain and the leads is
described by
\begin{eqnarray}
H_{\mathrm{T}}&=&\sum_{k,\alpha, i
}V_{k\alpha}[(\cos\frac{\theta_{\alpha}}{2}c_{k\alpha + }^{\dag}
-\sin\frac{\theta_{\alpha}}{2}c_{k\alpha-}^{\dag})d_{i\uparrow }\nonumber\\
&&\ \ \ \ \ \ \ \ \ +(\sin\frac{\theta_{\alpha}}{2}c_{k\alpha +
}^{\dag}+\cos \frac{\theta_{\alpha}}{2}c_{k\alpha
-}^{\dag})d_{i\downarrow}]+H.c.,\nonumber\\
\end{eqnarray}
where $\theta_{\alpha}$ is the angle between the easy axis of the
grain and the spin-quantization direction of lead $\alpha$. For
simplicity, we set the easy-axis of the grain as $z$-axis and assume
that $\theta_{(6)}=0$ and $\theta_{(4,5)}=\theta\in[0,\pi/2]$, as
shown in Fig. \ref{fig:axis}. In the following, we denote
$\mu_{+/-}^{(4)}=\mu_{\uparrow/\downarrow}^{(4)}$ when
$\theta_{(4,5)}=0$ for simplicity.

\subsection{\label{sec:grainGS}Ground branch $|N_0,S_0,S_z\rangle$ of the grain and possible excitations}

\begin{figure}[htbp]
\centering
\includegraphics[width=0.5\textwidth]{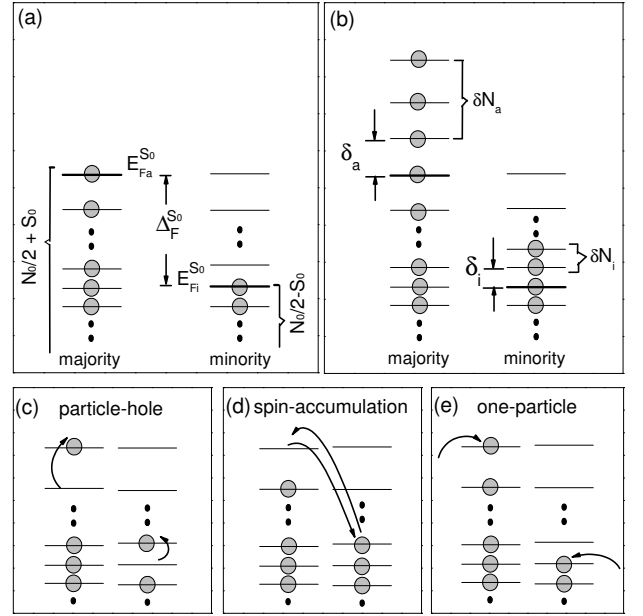}
\caption{The filling of one-particle levels in the majority and
minority bands of the grain with (a) $N_0$ electrons, and (b)
$N_0+\delta N$ electrons, where $\delta N=\delta N_a+\delta N_i$,
and $\delta N_{a}$ and $\delta N_{i}$ are the excess electrons added
to the majority and minority band, respectively. For the ground
branch with $N_0$ electrons in the grain, the magnitude of the
angular momentum $S=S_0$. $E_{Fa}^{S_0}$, and $E_{Fi}^{S_0}$ are the
highest occupied levels of the majority and minority bands for the
ground branch for $N=N_0$ and $S=S_0$.  $\delta_{a}$ and
$\delta_{i}$ is the level spacings near the $E_{Fa}^{S_0}$ and
$E_{Fi}^{S_0}$, respectively. (c) The particle-hole excitations,
with excitation energy $\sim \delta_i,\delta_a > 1$meV. (d) The
spin-accumulation excitations, with excitation energy $\sim J \sim
2$meV. These two excitations are high-energy excitations, with the
excitation energy larger than the spin bias $V_s$, thus can not be
excited and will be omitted. Note that according to Fig.
\ref{fig:Yang2008_gamma_jle_spinbias}, $V_s \sim 0.1 $ meV. (e)
One-particle excitations that changes the electron number by one. It
can add extra electrons to the majority or minority band. By tuning
the gate voltage, its excitation energy can be minimized down to
$\sim K_N\sim 0.01$meV, which is smaller than the spin bias $
V_s\sim 0.1$meV. Then, one-particle excitations is the only possible
excitation for the present device and parameters.
}\label{fig:filling}
\end{figure}

The eigen states of $H_G$ are labeled by $|\{n_i\},S,S_z\rangle$,
where $\{n_i\}$ denote the occupation on each level in the grain,
$S$ and $S_z$ are, respectively, the quantum numbers for the
magnitude of $\mathbf{\hat{S}}$ and $\hat{S}_z$. Because of the low
experimental temperature (as low as 20 mK) and the large Coulomb
repulsion $E_C$ ($> 30$ meV),\cite{Gueron1999.PRL.83.4148} the
charge fluctuation and particle-hole excitations [Fig.
\ref{fig:filling} (c)] are suppressed hence it is reasonable to
assume that the electrons in the grain compactly occupy all the
lowest available levels. These states are denoted as
$|N,S,S_z\rangle$, where $N=\sum_i n_i$ is the total electron number
in the grain.

When there is $N=N_0$ electrons in the grain, the competition
between the kinetic energy and the $J$ term will force $N_0/2+S_0$
electrons (the majority band) to orient anti-parallel with the rest
$N_0/2-S_0$ electrons (the minority band), leading to a nonzero
magnitude of angular momentum $S=S_0$ at the ground states(Stoner
instability), as shown in Figs. \ref{fig:axis} and \ref{fig:filling}
(a). This overall ground branch is denoted as $|N_0,S_0,S_z\rangle$,
where $S_z\in[-S_0,S_0]$. The $(2S_0+1)$-fold degeneracy of the
overall ground branch is lifted by the anisotropy, with two
degenerate ground states $|N_0,S_0,\pm S_0\rangle$.

There are three kinds of basic excitations from the ground branches
$|N_0,S_0,S_z\rangle$, as shown in Fig. \ref{fig:filling}(c), (d),
and (e). The particle-hole excitation destroys the compact
occupation of the one-particle levels, while does not change $N$,
$S$, and $S_z$. The spin-accumulation excitation changes $S$ by
moving electron between the majority and minority bands, while does
not change $N_0$ and $S_z$. The one-particle excitation changes $N$
by adding or removing electrons, which also leads to changes in $S$
and $S_z$.

Generally, the excited energy from $|N_0,S_0,S_z\rangle$ to
$|N_0+\delta N,S_0+\delta S,S_z+\delta S_z\rangle$ can be found with
the help of Figs. \ref{fig:filling}(a) and (b) as (using the
relations $\delta N=\delta N_a+\delta N_i$ and $\delta S=(\delta
N_a-\delta
N_i$)/2),\cite{Canali2000prl85_5623,Kleff2002prb65_214421}
\begin{eqnarray}\label{excitationenergy}
&&\delta E(\delta N, \delta S, \delta S_z)\nonumber\\
&\equiv& E_{|N_0+\delta N,S_0+\delta S,S_z+\delta
S_z\rangle}-E_{|N_0,S_0,S_z\rangle}\nonumber\\
&=&(\delta
N)^2[E_C+\frac{1}{8}\delta_{a}+\frac{1}{8}\delta_{i}]+\delta
N[\overline{E}_F^{S_0}+\frac{1}{4}\delta_{a}+\frac{1}{4}\delta_{i}+V_g]\nonumber\\
&+&(\delta
S)^2[\frac{\delta_{a}}{2}+\frac{\delta_{i}}{2}-\frac{J}{2}]+\delta S
[\frac{\delta_{a}}{2}-\frac{\delta_{i}}{2}+\Delta_F^{S_0}-J(S_0+\frac{1}{2})]\nonumber\\
&+&\delta N\delta S
[\frac{1}{2}\delta_{a}-\frac{1}{2}\delta_{i}]-\frac{K_N}{S_0}(2S_z\delta
S_z+\delta S_z^2),
\end{eqnarray}
where the definitions of parameters are given in Fig.
\ref{fig:filling} and
$\overline{E}_F^{S_0}=(E_{Fa}^{S_0}+E_{Fi}^{S_0})/2$.

\begin{table}[htpb]
\caption{Parameters for a $S_0=1000$ grain. All energies are in
meV.\cite{Canali2000prl85_5623,Kleff2002prb65_214421} $E_C$, $S_0$,
$K_N$ are estimated by
experiments,\cite{Gueron1999.PRL.83.4148,Deshmukh2001.PhysRevLett.87.226801,Jamet2001.prl.86.4676,Thirion2003.natmat.2.524}
$\delta_a$, $\delta_i$, $\Delta_F^{S_0}$ are from the band
calculations.\cite{Papaconstantopoulos1986book} According to Eq.
(\ref{excitationenergy}), $\overline{E}_F^{S_0}$ can be absorbed
into $V_g$, thus is set to 0 for convenience. $\mu^{(5)}(\infty)$ is
the energy zero point throughout the paper.} \label{tab:parameters}
\begin{ruledtabular}
\begin{tabular}{ccccccccccc}
   $E_C$&  $K_N$ & $\delta_{a}$ & $\delta_{i}$ & $\Delta_F^{S_0}$ & $\overline{E}_F^{S_0}$& $\mu^{(5)}(\infty)$\\
  \hline
   30 & 0.01   & 4.61  & 1.19 & 2000 & 0 & 0\\
\end{tabular}
\end{ruledtabular}
\end{table}
We will adopt a set of parameters for a grain with $S_0=1000$, as
given in Tab. \ref{tab:parameters}. The theoretical
calculations\cite{Canali2000prl85_5623,Kleff2001prb64_220401,Kleff2002prb65_214421}
based on these parameters are consistent with most features observed
in the experimental transport
spectra.\cite{Gueron1999.PRL.83.4148,Deshmukh2001.PhysRevLett.87.226801}
With these parameters, the value of $J$ can be deduced using a
saddle point analysis as follows.

\subsection{\label{sec:saddlepoint}Range of $J$ for a stable $S_0$ by saddle-point analysis}

Above we just assume that $S=S_0$ when there are $N=N_0$ electrons
in the grain. Since $S_0$ originates from the competition between
the $J$ term and the kinetic energy, its value should be calculated
for the given values of $J$ and $\delta_{a,i}$. However, since we
already know $S_0\sim 1000 $ from the
experiments,\cite{Gueron1999.PRL.83.4148,Deshmukh2001.PhysRevLett.87.226801,Jamet2001.prl.86.4676,Thirion2003.natmat.2.524}
and $\delta_{a,i}$ from the band
calculation,\cite{Papaconstantopoulos1986book} we will use these
data to conversely deduce the value of $J$, then generate other
information (e.g., one-particle excitations) at the proximity of the
deduced $J$, much like a saddle-point analysis.

The stability of the branch $|N=N_0,S=S_0\rangle$ requires that its
energy should be at least locally minimal compared to the states
$|N=N_0,S=S_0\pm 1\rangle$, i.e.,
\begin{eqnarray}\label{stablecondition}
\delta E(\delta N=0, \delta S=+ 1)
&=&\delta_{a}+\Delta_F^{S_0}-J(S_0+1)\geq 0,\nonumber\\
\delta E(\delta N=0, \delta S=- 1) &=&
\delta_{i}-\Delta_F^{S_0}+JS_0 \geq 0,
\end{eqnarray}
where we have used Eq. (\ref{excitationenergy}) and ignored the
anisotropy term because $K_N$ is much smaller than other energies.
In other words, Eq. (\ref{stablecondition}) leads to that, for $J$
belongs to the range
\begin{eqnarray}\label{stability of overall ground state}
\frac{\Delta_F^{S_0}-\delta_{i}}{S_0} \leq J \leq
\frac{\Delta_F^{S_0} +\delta_{a}}{S_0+1},
\end{eqnarray}
the grain will adopt a stable $S=S_0$ when there are $N_0$ electrons
in the grain. Once $J$ exceeds this range, the overall ground branch
will evolve to adopt a smaller or larger $S=S_0'=S_0\mp1$, then Eq.
(\ref{stability of overall ground state}) still holds for the new
$S_0'$ (note that $\Delta_F^{S_0\pm 1}=\Delta_F^{S_0}\pm \delta_a\mp
\delta_i$).

\subsection{\label{sec:oneparticleexcitation}One-particle excitations to the branches $|N_0+1,S_0\pm 1/2,S_z\rangle$}

\begin{figure}[htbp]
\centering
\includegraphics[width=0.5\textwidth]{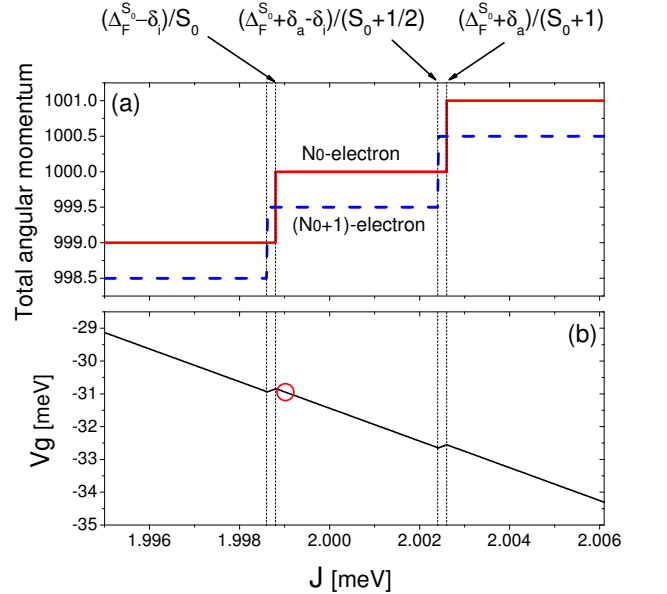}
\caption{(a) Total angular momenta for the ground branches of the
grain with $N_0$ and $N_0+1$ electrons, respectively. (b) Along the
zigzag line of $(J,V_g)$, the $N_0$- and $(N_0+1)$-electron ground
branches are nearly degenerate. The parameters are listed in Table
\ref{tab:parameters}.}\label{fig:phasediagram}
\end{figure}

The excitation energies for the particle-hole excitation shown in
Fig. \ref{fig:filling} (c) and spin-accumulation excitation shown in
Fig. \ref{fig:filling} (d) are of the order of $\delta_{a/i}$ and
$J$,
respectively.\cite{Canali2000prl85_5623,Kleff2002prb65_214421,Parcollet2006.prb.73.144420}
According to the parameters for $S_0=1000$ shown in Tab.
\ref{tab:parameters}, these excitations are of meV, much larger than
the $V_s \sim 0.05$ meV estimated in Fig.
\ref{fig:Yang2008_gamma_jle_spinbias}, thus will be excluded. In the
following, we will only consider the one-particle excitation from
the $N_0$ to $N_0+1$ electrons, as shown by Fig. \ref{fig:filling}
(e), i.e., adding excess electrons to the grain.

Again, because $E_C \gg k_B T$, the gate voltage in this work is
restricted so that one and only one excess electron ($\delta N=1$)
can be added into the grain. Depending on this excess electron being
added to the majority or minority band, the magnitude of the angular
momentum could change to $S=S_0+1/2$ or $S_0-1/2$, respectively.
With the help of Eqs. (\ref{excitationenergy}) and (\ref{stability
of overall ground state}), we find that, when
\begin{eqnarray}\label{minrange}
\frac{\Delta_F^{S_0}-\delta_{i}}{S_0}\leq J \leq
\frac{\Delta_F^{S_0}+\delta_{a}-\delta_{i}}{S_0+1/2},
\end{eqnarray}
the $N_0+1$-electron ground branch will adopt $S=S_0-1/2$, and the
required transition energies from $|N_0,S_0,S_z\rangle$ are
\begin{eqnarray}\label{chargingenergymin}
&& E_{|N_0+1,S_0- \frac{1}{2},S_z\pm
\frac{1}{2}\rangle}-E_{|N_0,S_0,S_z\rangle} \nonumber\\&=&
\delta_{i}+V_g+E_C+E_{Fi}^{S_0}
+\frac{J}{2}(S_0+\frac{1}{4})-\frac{K_N}{S_0}(\frac{1}{4}\pm
S_z).\nonumber\\
\end{eqnarray}
On the contrary, when
\begin{eqnarray}\label{majrange}
\frac{\Delta_F^{S_0}+\delta_{a}-\delta_{i}}{S_0+1/2}\leq J\leq
\frac{\Delta_F^{S_0} +\delta_{a}}{S_0+1},
\end{eqnarray}
the $N_0+1$-electron ground branch will adopt $S=S_0+1/2$, and the
required transition energies from $|N_0,S_0,S_z\rangle$ are
\begin{eqnarray}\label{chargingenergymaj}
&& E_{|N_0+1,S_0+ \frac{1}{2},S_z\pm
\frac{1}{2}\rangle}-E_{|N_0,S_0,S_z\rangle}\nonumber\\
&=& \delta_{a}+V_g+E_C+E_{Fa}^{S_0}
-\frac{J}{2}(S_0+\frac{3}{4})-\frac{K_N}{S_0}(\frac{1}{4}\pm S_z). \nonumber\\
\end{eqnarray}

To summarize the above saddle-point analysis, the angular momenta of
$N_0$- and $(N_0+1)$-electron ground branches as a function of $J$
are shown in Fig. \ref{fig:phasediagram}(a), in which the arrows
mark the ranges indicated by Eqs. (\ref{stability of overall ground
state}), (\ref{minrange}) and (\ref{majrange}).

According to Fig. \ref{fig:phasediagram}(a), for most value of $J$,
$S$ of the $(N_0+1)$-electron branch is $1/2$ smaller than that of
the $N_0$-electron branch. This is a direct results of
$\delta_i<\delta_a$. Although $J$ is not a tunable quantity, Fig.
\ref{fig:phasediagram}(a) implies that in reality the excess
electron is far more likely to occupy the minority band than the
majority band, which is also consistent with the previous
literatures.\cite{Canali2000prl85_5623,Waintal2005prl94_247206}
Therefore, in the following we will mainly consider the one-particle
excitations between the ground branches $|N_0,S_0,S_z\rangle$ and
$|N_0+1,S_0-1/2,S_z\rangle$.

\subsection{\label{sec:stateforsimulation}States used for numerical simulations}

Remember we have set $\mu^{(5)}(\infty)$ as the energy zero point.
With respect to $\mu^{(5)}(\infty)$, we can always choose suitable
$V_g$ in Eqs. (\ref{chargingenergymin}) and
(\ref{chargingenergymaj}) to compensate $E_C$ and other energies, so
that the ground-branches with $N_0$ and $N_0+1$ electrons can be
tuned to be nearly degenerate. These $V_g$ as a function of $J$ are
shown in Fig. \ref{fig:phasediagram}(b). For instance, in Eq.
(\ref{chargingenergymin}), by choosing
$V_g=-\delta_{i}-E_C-E_{Fi}^{S_0} -J(S_0+1/4)/2+K_N/4S_0$, one
obtains
\begin{eqnarray}\label{chargingenergy_Vg}
E_{|N_0+1,S_0- \frac{1}{2},S_z\pm
\frac{1}{2}\rangle}-E_{|N_0,S_0,S_z\rangle}=\mp\frac{K_N}{S_0}S_z.
\end{eqnarray}
In this context, the energy required to add an electron from lead
(4,5) into the grain is related to only the magnetization of the
grain. According to Eq. (\ref{chargingenergy_Vg}), the spectrum
width of $E_{|N_0+1,S_0- \frac{1}{2},S_z\pm
\frac{1}{2}\rangle}-E_{|N_0,S_0,S_z\rangle}$ is $2K_N$ for all the
possible $S_z\in [-S_0, S_0]$. This value is the key to determine
the critical $I_c$ in the Sec. \ref{sec:criticalIc}.

Experimentally, it is easy to find the suitable $V_g$ at which the
ground-branches with $N_0$ and $N_0+1$ electrons are
nearly-degenerate, as in the usual transport
experiments.\cite{Gueron1999.PRL.83.4148,Deshmukh2001.PhysRevLett.87.226801}
For example, one just apply a small charge bias voltage $> 2K_N$
between lead (4,5) and lead (6), and measure the current through the
grain while scanning $V_g$, like a usual source-gate-drain
measurement. Because of the Coulomb blockade, the grain can not
conduct electrons unless the $N_0$- and $(N_0+1)$-electron
ground-branches are degenerate. Therefore, the nearly-degenerate
situation is find as: at which $V_g$, the grain is conducting under
a small charge bias voltage between (4) and (6).

Based on the above analysis and discussions from Appendixes
\ref{sec:grainGS}-\ref{sec:stateforsimulation}, in the following, we
will consider mainly the one-particle excitations from the branches
$|N_0, S_0,S_z\rangle$ to $|N_0+1, S_0-1/2,S_z\rangle$ and when
their energies are nearly degenerate. Specifically, we will choose a
set of $(J,V_g)$ marked by the circle in Fig.
\ref{fig:phasediagram}(b) for our numerical simulations, where
$J=\Delta_F^{S_0}/(S_0+1/2)$, and $V_g=-\delta_{i}-E_C-E_{Fi}^{S_0}
-J(S_0+1/4)/2+K_N/4S_0+\Delta V_g$, $\Delta V_g $ is a small
variation of the gate voltage that drives the grain away from the
nearly-degenerate point of $N_0$- and $(N_0+1)$-electron ground
branches.

\section{\label{sec:ValidityRateEquation}Validity of Rate
equations}

Although the rate equations formalism is widely employed for the
mesoscopic systems weakly coupled to the electrodes, its validity
deserves some
discussion.\cite{Beenakker1991.PRB.44.1646,vonDelft2001.PhysRep.345.61}

In the previous works by Waintal \emph{et
al},\cite{Waintal2005prl94_247206,Parcollet2006.prb.73.144420} the
intrinsic spin relaxation is considered in terms of coupling to a
bosonic bath. We do not consider this effect for two reasons: (i)
According to Eq. (3) of Ref. \onlinecite{Waintal2005prl94_247206},
the intrinsic relaxation will lead to a term similar to the Gilbert
damping, which tends to relax the grain to one of its two degenerate
maximally magnetized ground states, e.g., $|N_0,S_0,-S_0\rangle$ or
$|N_0,S_0,S_0\rangle$. In the following, we will mainly discuss how
to use the NLSV signal to read out the magnetization of the grain
(Sec. \ref{sec:nlsv}), which already limits the discussion to these
maximally magnetized states. Therefore, the results in Sec.
\ref{sec:nlsv} will be qualitatively unaffected by the intrinsic
relaxation. (ii) For the spin bias -induced magnetization switching
discussed in Sec. \ref{sec:switching}, the results are only valid
when the coupling between the grain and the lead is dominant and
much smaller in time scale than the intrinsic relaxation. On the
other hand, the Born and Markoff approximations enforce that
$\Gamma$ is much smaller than $k_B T$ (bath temperature) and the
energy difference between many-body states in the grain. These two
requirements confine the validity range of $\Gamma$ in this work.

Besides, the particle-hole excitation terms in Eq. (10) of Ref.
\onlinecite{Kleff2002prb65_214421} are also absent in this work,
because the energy scale of these excitations are of the order
$\delta_{a},\delta_i >$ 1meV, which already one order larger than
the spin bias ($V_s\sim 0.1$meV) that can be generated according to
Fig. \ref{fig:Yang2008_gamma_jle_spinbias}.


\end{document}